 \documentclass[final,3p,times,twocolumn,sort&compress, lefttitle]{elsarticle}

\usepackage{amssymb}
\usepackage{amsmath}
\usepackage{booktabs}
\usepackage{gensymb}
\usepackage{booktabs}
\usepackage{caption}
\usepackage{subcaption}
\usepackage{multirow}
\usepackage{xcolor}
\usepackage{hyperref}

\journal{Acta Astronautica}

\newcommand{\rev}[1]{\textcolor{red}{#1}}
\newenvironment{revblock}{\color{red}}{\color{black}}

\renewcommand{\rev}[1]{#1}  
\renewenvironment{rev}{}{}  
\newcommand{\brev}{\begin{revblock}}
\newcommand{\erev}{\end{revblock}}

\begin{document}

\begin{frontmatter}



\title{Design of a rapid transit to Mars mission using laser-thermal propulsion}

%
%

\author{Emmanuel Duplay\corref{cor1}}
\ead{emmanuel.duplay@mail.mcgill.ca}
\author{Zhuo Fan Bao}
\author{Sebastian Rodriguez Rosero}
\author{Arnab Sinha}
\author{Andrew Higgins}
\ead{andrew.higgins@mcgill.ca}

\cortext[cor1]{Corresponding author}

\address{Department of Mechanical Engineering, McGill University, Montreal, Quebec, H3A 0C3, Canada}


\begin{abstract}
    The application of directed energy to spacecraft mission design is explored using rapid transit to Mars as the design objective. An Earth-based laser array of unprecedented size (10~m diameter) and power (100~MW) is assumed to be enabled by ongoing developments in photonic laser technology. A phased-array laser of this size and incorporating atmospheric compensation would be able to deliver laser power to spacecraft in cislunar space, where the incident laser is focused into a hydrogen heating chamber via an inflatable reflector. The hydrogen propellant is then exhausted through a nozzle to realize specific impulses of 3000 s. The architecture is shown to be immediately reusable via a burn-back maneuver to return the propulsion unit while still within range of the Earth-based laser. The ability to tolerate much greater laser fluxes enables realizing the combination of high thrust and high specific impulse, making this approach favorable in comparison to laser-electric propulsion and occupying a parameter space similar to gas-core nuclear thermal rockets (without the requisite reactor). The heating chamber and its associated regenerative cooling and propellant handling systems are crucial elements of the design that receive special attention in this study. The astrodynamics and the extreme aerocapture maneuver required at Mars arrival after a 45-day transit are also analyzed in detail. The application of laser-thermal propulsion as an enabling technology for other rapid transit missions in the solar system and beyond is discussed.
\end{abstract}


%

\end{frontmatter}
    \raggedbottom
    \section{Introduction}
        Recent developments in photonics—in particular, the emergence of inexpensive fiber-optic laser amplifiers—have revitalized interest in directed-energy propulsion. The ability to phase-lock large arrays of fiber-optic laser amplifiers together in a modular fashion, enabling them to operate as a single optical element of arbitrarily large size and power, has now been demonstrated at laboratory scales \cite{srinivasan_directed_2019-1,srinivasan_directed_2019}. The application of atmospheric compensation techniques originally developed for astronomy (i.e., adaptive optics that can effectively remove the beam distortions caused by Earth’s atmosphere) would allow the laser array to be built on earth as opposed to in space \cite{bandutunga_photonic_2021,hettel_beam_2021}. Past work has been done on applying these developments to interstellar flight. Indeed, dense laser arrays on the scale of kilometers with fluxes on the order of 1~kW/m$^2$ leaving the array would enable true interstellar missions, wherein the photon pressure of the laser would quickly propel 1-m-scale lightsails to 20--30\% of lightspeed. If directed toward nearby solar systems, such a lightsail could return images from neighboring exoplanets within a 25-year mission \cite{ lubin_roadmap_2016,parkin_breakthrough_2018}.
        
        More near-term applications of directed energy for interplanetary flight are better suited to using a reaction mass to couple the delivered laser energy to change the momentum of the spacecraft. Options using laser-based directed energy are laser-electric propulsion \cite{brophy_directed-energy_2019,sheerin_fast_2021} and laser-thermal propulsion (LTP\footnote{AU = Astronomical Unit, CME = Coronal Mass Ejection, GCNR = Gas-Core Nuclear Rocket, GCR = Galactic Cosmic Rays, LE = Laser Electric, LEP = Laser-Electric Propulsion, LSC = Laser-Supported Combustion, LTP = Laser-Thermal Propulsion, LTPS = Laser-Thermal-Propulsion System, NTR = Nuclear Thermal Rocket, PPS = Power and Propulsion System, PV = photovoltaic}). Laser-thermal propulsion is further classified into (1) laser ablation propulsion using an initially condensed-phase reaction mass \cite{phipps_review_2010} and (2) laser-thermal propulsion with a gaseous propellant (typically hydrogen) that is heated and expanded through a nozzle. The second approach is well-matched to the continuous-wave nature of phased-array lasers employing atmospheric correction. The application of this mode of laser-thermal propulsion using a large phased-array laser to deep-space mission design is the subject of the present paper.
        
        Laser-thermal propulsion was extensively studied starting in the 1970s when the first continuously burning hot plasma sustained by a laser was observed under laboratory conditions as reported in \cite{generalov_continuous_1970}. This discovery was soon followed by a speech by Arthur Kantrowitz \cite{kantrowitz_relevance_1971} suggesting the use of lasers to directly heat propellant within a rocket, springboarding the development of LTP for the following three decades. Despite promising preliminary results, according to a historical record \cite{jones_brief_2003} of laser propulsion research at the Marshall Space Flight Center, the lack of funds to maintain the complex laser systems and the lack of political interest contributed to the termination of experimental research being conducted in this field in the mid-1980s.
        
        Studies from this period usually considered gasdynamic CO$_2$ lasers operating at a 10.6-µm-wavelength, the most powerful lasers at the time. This longer wavelength and the meter-scale monolithic optics then available limited consideration of laser-thermal propulsion to orbit transfer in near-Earth space applications \cite{caveny_orbit-raising_1984}. The transition to the 1-µm operating wavelength of present-day fiber-optic lasers and the ability to combine them into a massively parallel, phased array of large effective optical diameter means that the focal length over which the laser can deliver energy (i.e.,\footnote{See Nomenclature at end of paper} $d_\mathrm{f} \sim D_\mathrm{e} D_\mathrm{r} / \lambda$) can be extended by two orders of magnitude or more, making the application of laser-thermal propulsion for deep-space missions of interest.
        Thus, a revisit to mission design applications of laser-thermal propulsion is warranted.
        
        A recent NASA solicitation seeking revolutionary propulsion for rapid, deep-space transit identified a number of candidate missions of interest: traversing the distance between Earth orbit and Mars orbit in no more than 45 days, traversing a distance of 5~AU in no more than one year, traversing a distance of 40 AU in no more than five years, and traversing a distance of 125 AU in no more than ten years \cite{national_aeronautics_and_space_administration_space_2018}. The Mars-in-45-day requirement is presumably motivated by concern over astronaut exposure to galactic cosmic rays (GCRs) and the potential threat of coronal mass ejections (CMEs) in transit. Recent in-situ measurements by the NASA Curiosity Rover have shown that the radiation environment on the surface of Mars is a factor of two lower than that experienced in transit to Mars once outside the Earth’s protective magnetosphere \cite{berger_long_2020}. This finding suggests an emphasis should be placed on propulsion technologies coupled to mission architectures that minimize the transit time to Mars for crewed missions.
        
        In a larger sense, rapid missions to Mars have become a convenient metric in comparing different propulsion technologies \cite{pelaccio_examination_2002}, including nuclear thermal \cite{borowski_nuclear_2012}, nuclear electric \cite{chang-diaz_fast_2013,berend_feasibility_2014}, solar electric \cite{chang-diaz_solar_2019}, and other high specific impulse and high specific power technologies such as fusion \cite{genta_achieving_2020}. For this reason, we have selected a rapid-transit-to-Mars mission as the baseline design for this study.
        
        In this paper, the implications of using a 10-m-scale phased-array laser based on Earth and applied to rapid transit missions within the solar system and rapid transits to Mars in particular are explored. The use of large inflatable reflectors with high reflectivity and the ability to tolerate intense laser flux to focus the laser power delivered to the spacecraft into a hydrogen heating chamber is able to generate specific impulses and thrust-levels (upon expansion through a nozzle) comparable to advanced gas-core nuclear thermal rockets ($I_\mathrm{sp} \approx$ 2500--3000~s).
        The design of the heating chamber is identified as the crucial element of the architecture and is explored in detail in the present study. The propellant storage, regenerative cooling, and delivery system are also considered. Optimization of the transfer orbit utilizing the large $\Delta v$ available in near-Earth space (within the range of the 10-m-scale laser) is conducted. Since a laser will not be available for deceleration upon arrival at Mars, considerable attention is also focused on advanced aerocapture techniques necessitated by the large approach velocities. The capability of having the Mars-injection propulsion stage to return to Earth by effectuating a burn-back maneuver while still within the focal range of the Earth-based laser is studied in comparison to a one-off-use scenario. Trade-offs between lower $\Delta v$ and greater payload fraction missions (e.g., for cargo delivery) will also be explored. The specific mission requirements for the design study reported here are derived from the NASA solicitation discussed above, however, this architecture could be used for a number of missions, including missions to the ice giant planets ($<$ 5 year transit time), missions into the interstellar medium at 125 AU ($<$ 10 year transit time), and missions to the solar gravitational focus starting at 550 AU ($<$ 50 year transit time) \cite{turyshev_direct_2020}.
        \begin{figure*}[!ht]
            \centering
            \includegraphics[width=\textwidth]{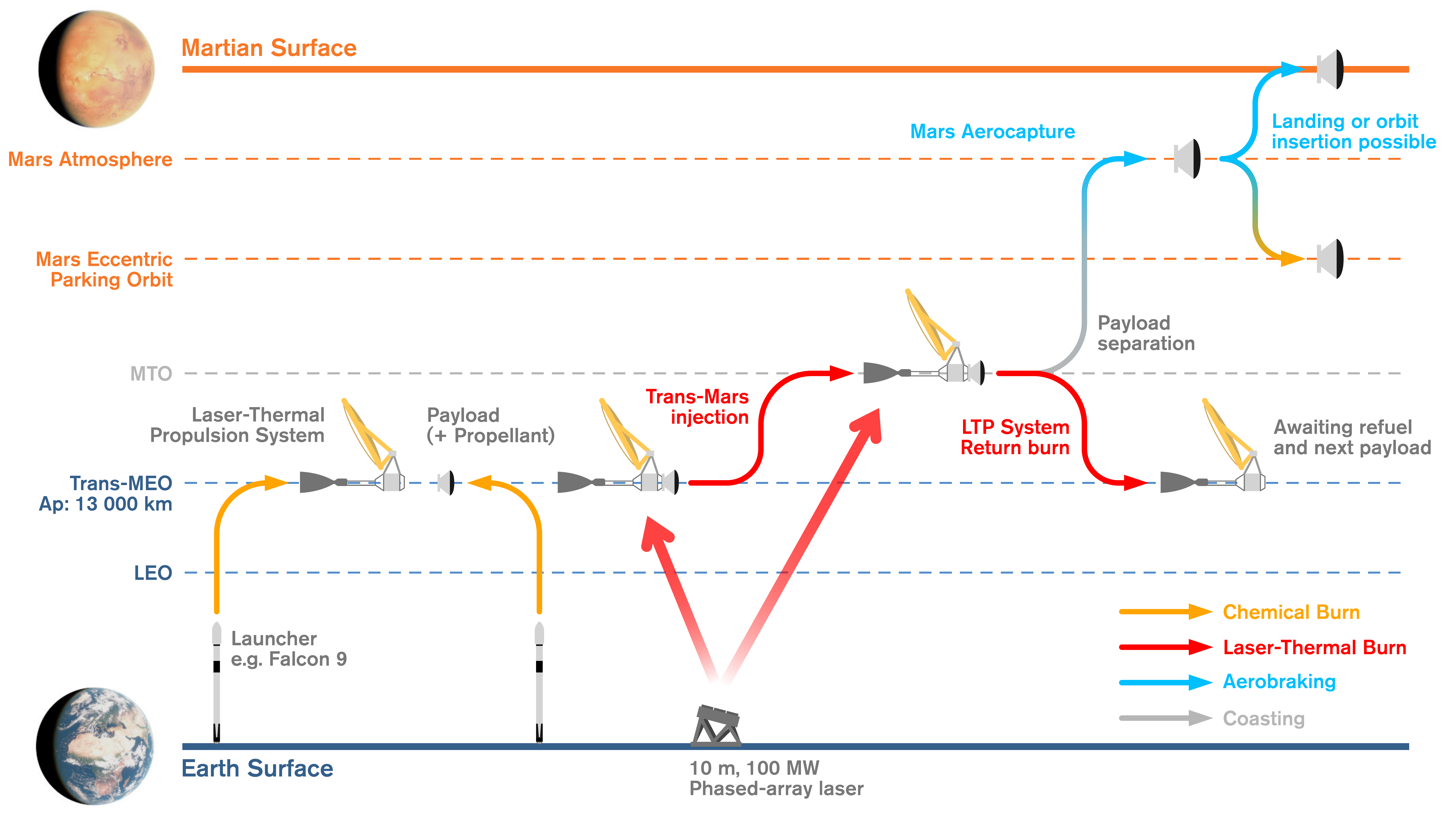}
            \caption{Concept of Operation diagram for a reusable Laser-Thermal Propulsion System}
            \label{fig:conops}
        \end{figure*}

    \section{Rapid Mars Transit Mission Design}
        \label{sec:mars1}
        \subsection{Architecture}
            An LTP system for interplanetary transfers would require a 10-m-diameter laser array operating at up to 100~MW, allowing the array to focus on a target up to 50~000~km away. LTP maneuver durations for missions considered in this study would range from several minutes to an hour (depending on the mission), eliminating the need to build several arrays around the planet or in orbit to ensure the continuous supply of laser power, as would be the case for laser electric propulsion. This feature of laser-thermal propulsion makes this proposed architecture an attractive application to early prototype laser arrays, serving as a stepping stone toward the more ambitious infrastructure (kilometer-scale arrays) required for directed-energy interstellar travel.
        
            \begin{figure}[t]
                \centering
                \includegraphics[width=\columnwidth]{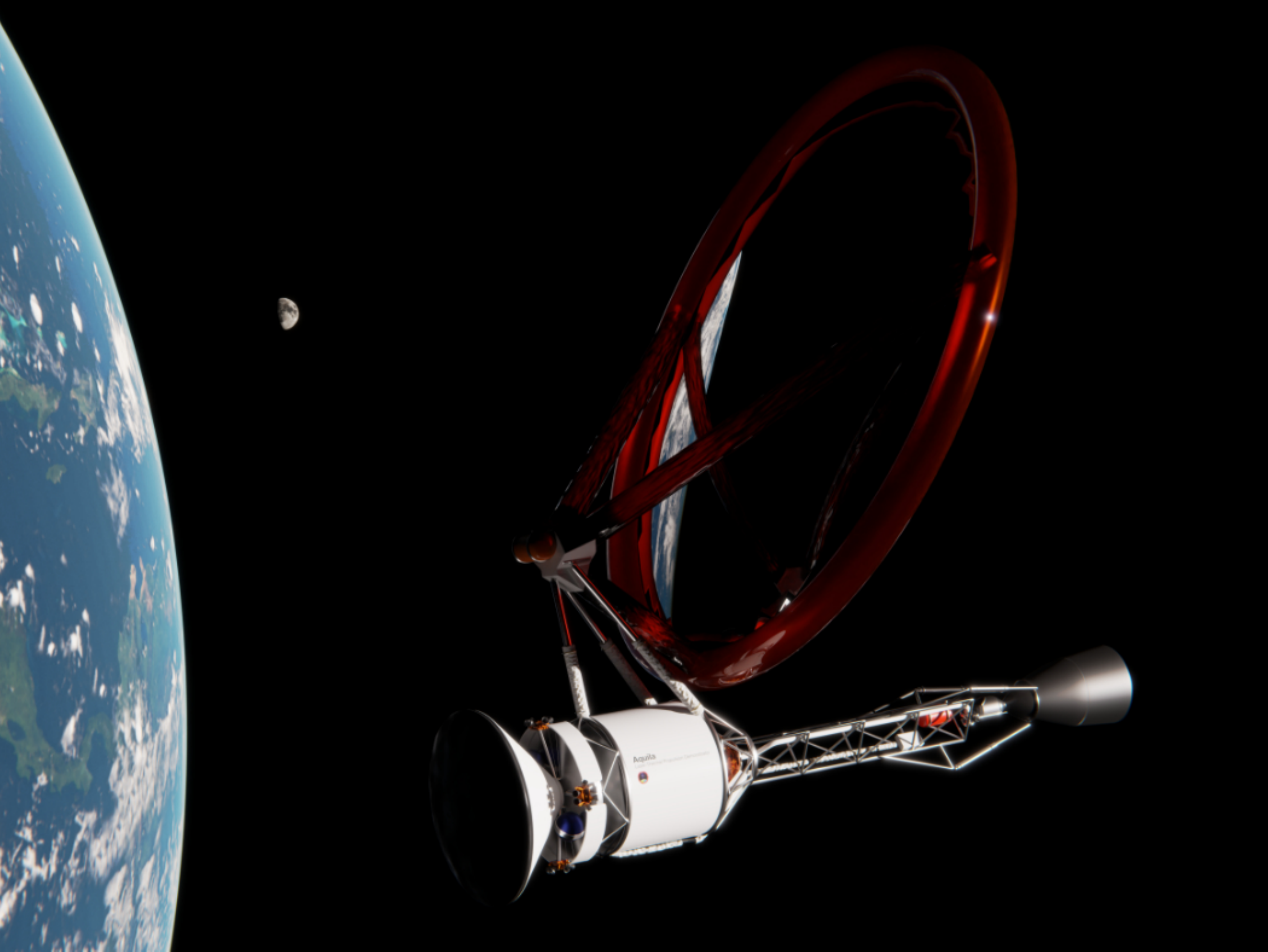}
                \caption{Conceptual render of a Laser-Thermal-Propulsion System, carrying a 1~ton payload for a 45-day Mars transfer}
                \label{fig:ltpRender}
            \end{figure}
            
        \subsection{Mission Profile}
            The mission concept of operations is presented in Figure \ref{fig:conops}. The mission begins with the launch of the laser-thermal-propulsion system (LTPS), hydrogen propellant, and payload, either separately or as a single launch of a Falcon 9 or Atlas-class launch vehicle. The fueled LTPS and payload, depicted in Figure \ref{fig:ltpRender}, are placed in an elliptical orbit with apogee above the van Allen belts (Trans-MEO, 13000$\times$500~km). This orbit permits the required dwell time over the ground-based laser such that the laser-powered propulsive maneuver---with a duration on the order of one hour---can occur during a single pass over the laser ground site. At the end of this propulsive maneuver, the payload is released into the high-energy transfer orbit to Mars. Following payload release, the LTPS---still in view and within the focal range of the ground-based laser---performs a second laser-powered maneuver to return it to the original elliptical orbit; this enables the entire LTPS hardware to be quickly reused following on-orbit propellant transfer. The payload requires only minor propulsive corrections to the trajectory during the short-duration ballistic transfer to Mars, followed by aerocapture upon arrival to Mars. Given the large approach velocity upon arrival at Mars, direct entry is not deemed feasible, however, aerocapture can be followed either by entry and landing or by insertion into a parking orbit around Mars.
            
            \begin{figure}[ht]
                \centering
                \includegraphics[width=\columnwidth]{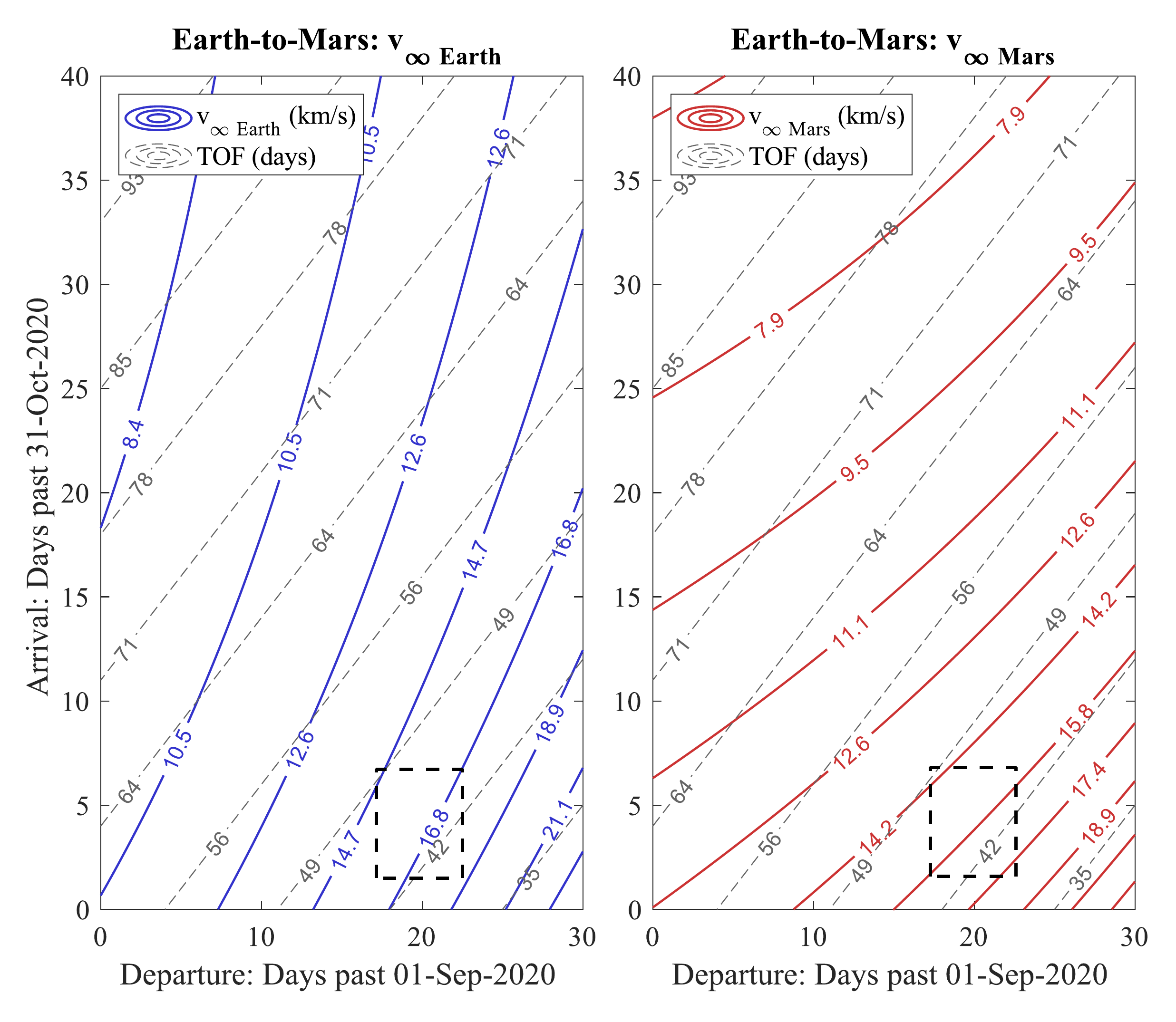}
                \caption{Porkchop curves of hyperbolic excess velocity for Earth departure (left) and arrival at Mars (right). Targeted region for a fast Mars transfer is identified with a dashed rectangle.}
                \label{fig:porkchop}
            \end{figure}
            
        \subsection{Astrodynamics}
            To ensure that the spacecraft remains within focal range and line-of-sight of the laser array throughout the propulsive maneuver, its trajectory was simulated assuming the start of the maneuver 30~minutes before reaching apogee. Given the propellant mass calculated in Table \ref{tab:altMissions} (line 1, propellant mass $m_\text{pr} = 700$ kg) for a single-use mission and a constant mass flow rate of 0.2~kg/s, the maneuver is expected to last 58~minutes. Accounting for the rotation of the Earth, the angle swept by the spacecraft in the sky relative to a ground observer is estimated to be 60°, with the maneuver ending before reaching geostationary altitude, well within the range of the 10-m-diameter laser array.
            
\begin{figure}[t]
                \centering
                \includegraphics[width=\columnwidth]{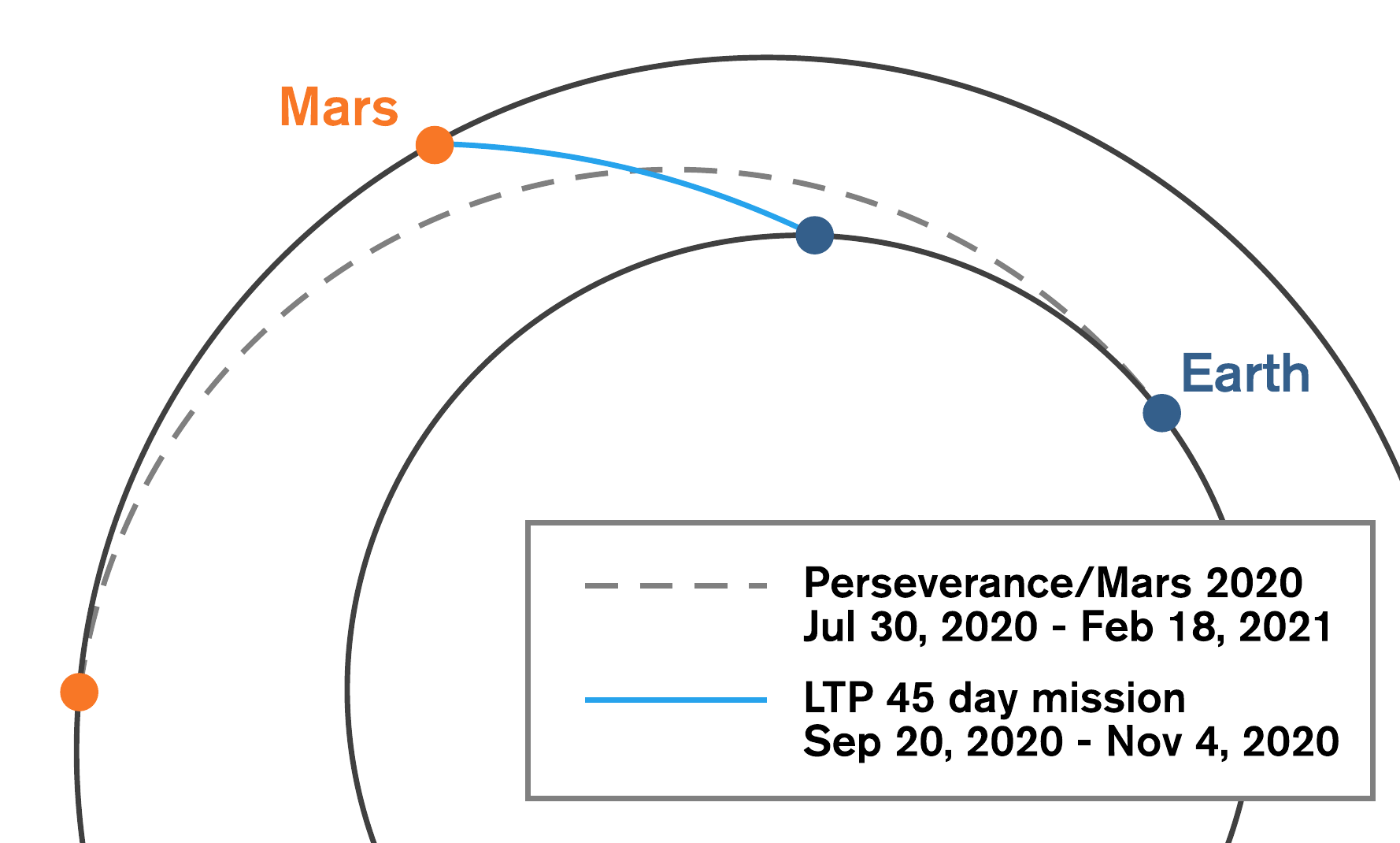}
                \caption{Mars transfer trajectory comparison. Orbits are to scale.}
                \label{fig:transferPlot}
            \end{figure}

            A custom MATLAB implementation of the algorithms presented in \cite{curtis_orbital_2005} to solve Lambert's problem was used to determine the precise $\Delta v$ requirements of this 45-day transfer during a 2020 launch window, in order to compare the mission time to recently launched Mars missions. Astrodynamic solutions for this launch window are plotted in Figure \ref{fig:porkchop}. An optimal 45-day transfer is depicted in Figure \ref{fig:transferPlot} for comparison to the Perseverance rover mission launching a month and a half earlier yet arriving three and a half months later than an LTP-launched payload. This particular trajectory, which would have launched on 20 Sept. 2020, required 13.95~km/s of $\Delta v$ from the LTPS parking orbit, and serves as a performance target for our vehicle design.
            
            \begin{table}[b]
                \centering
                \caption{45-day Mars transfer parameters}
                \label{tab:mars1param}
                \begin{tabular}{@{}lrl@{}}
                    \toprule
                    Parameter                                & Value & Unit \\ \midrule
                    Departure orbit apogee       & 13~000                     & km   \\
                    Final Martian orbit altitude    & 150                       & km   \\
                    Time of flight        & 45                        & days \\
                    Earth $v_\infty$        & 16.00                 & km/s \\
                    Mars $v_\infty$         & 15.41                & km/s \\
                    Earth departure $\Delta v$  & 13.95                 & km/s \\
                    Mars capture $\Delta v$ & 12.65                 & km/s \\ \bottomrule
                \end{tabular}
            \end{table}
            
        \subsection{Arrival}
            Although the performance targets presented in NASA's solicitation for ``Revolutionary Propulsion for Rapid Deep Space Transit" could have been satisfied with a flyby of Mars within 45 days, the value of such a mission is limited. This study therefore considered the feasibility of capturing the payload in Martian orbit, despite the lack of a laser array on Mars to provide the necessary $\Delta v$. Capture at Mars is a considerable challenge: As can be seen in Table \ref{tab:mars1param}, the $\Delta v$ required to insert the vehicle in a 150~km orbit about Mars is comparable to that of departure. Performing such a maneuver with chemical propellant (assumed $I_\mathrm{sp} = 451$~s) is not feasible, as this would reduce the useful payload mass to less than 6\% of the original 1000~kg. Without a laser-array at our destination, the only other way to decelerate is to perform an aerocapture maneuver.
            
            \subsubsection{Aerocapture Modeling}
\begin{figure}[b]
                    \centering
                    \includegraphics[width=\columnwidth]{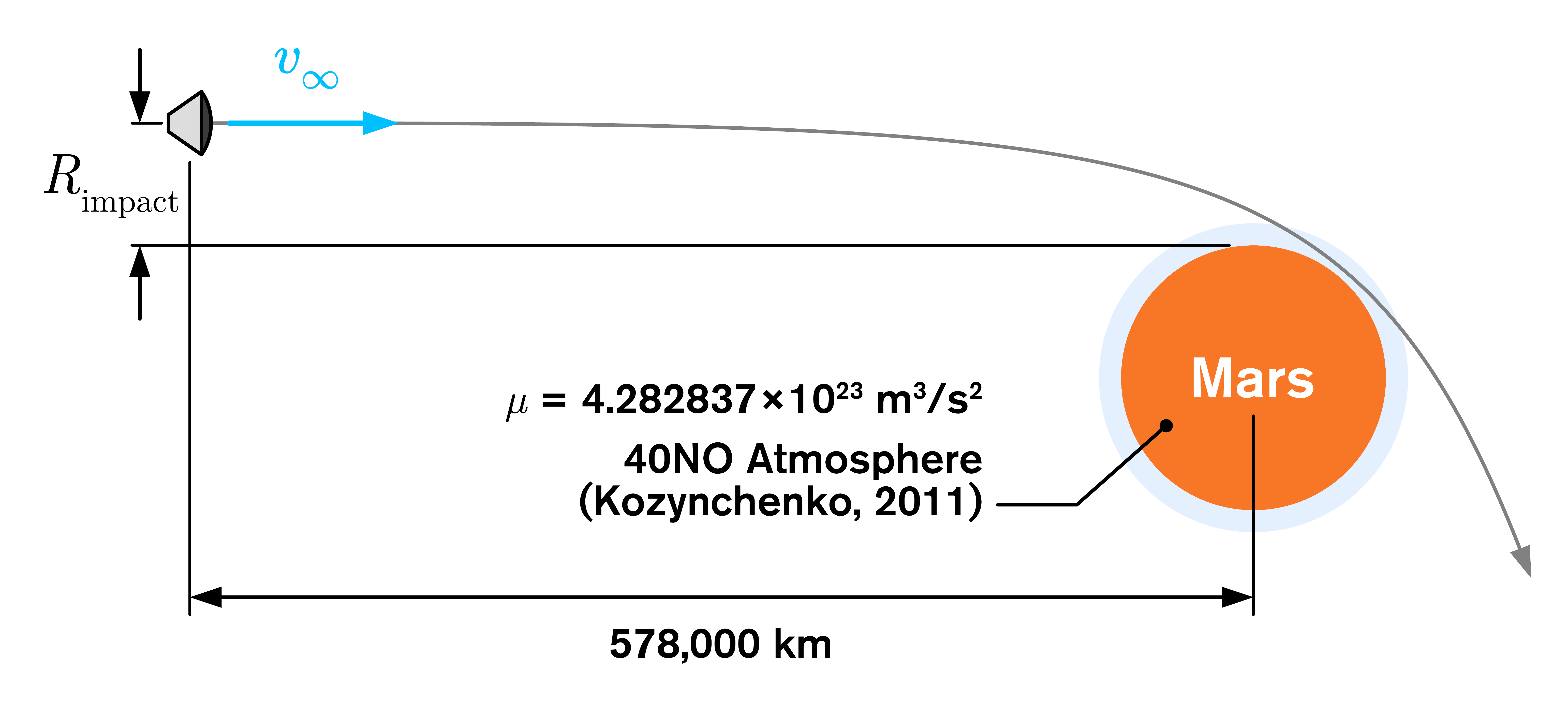}
                    \caption{2D numerical aerocapture model setup}
                    \label{fig:aerocaptureNumModel}
                \end{figure}

                A simple two-dimensional (2D) numerical aerocapture model was implemented to search for viable trajectories that would leave the spacecraft in an elliptical orbit once it exits the atmosphere, without imparting excessive thermal or acceleration loads. As shown in Figure \ref{fig:aerocaptureNumModel}, the entry vehicle is placed at a given aiming radius $R_\mathrm{Mars} + R_\mathrm{impact}$ and allowed to dive into the atmosphere to dissipate energy. The approximation of the 40NO Martian atmospheric model by Kozynchenko \cite{kozynchenko_analysis_2011} was deemed sufficient given its ease of implementation. The effects of gravity, drag, and lift were validated against classical orbital dynamics and 2D simulations of atmospheric re-entry presented in \cite{kluever_space_2018}. Thermal loads were modeled by computing both the stagnation point convective \cite{sutton_general_1971} and radiative \cite{tauber_stagnation-point_1991} heating rates, assuming a nose radius of 3.5~m. When possible, known constants for Mars were used in calculating the radiative heat flux. As no data for the radiative heating velocity function (tabulated in \cite{tauber_stagnation-point_1991}) was available for high velocity entries in the Martian atmosphere, values for Earth were used instead as radiative heating rates on either planet become comparable at velocities greater than 10~km/s \cite{lohar_optimal_1995}.
                
                \begin{figure}[b]
                    \centering
                    \includegraphics[width=\columnwidth]{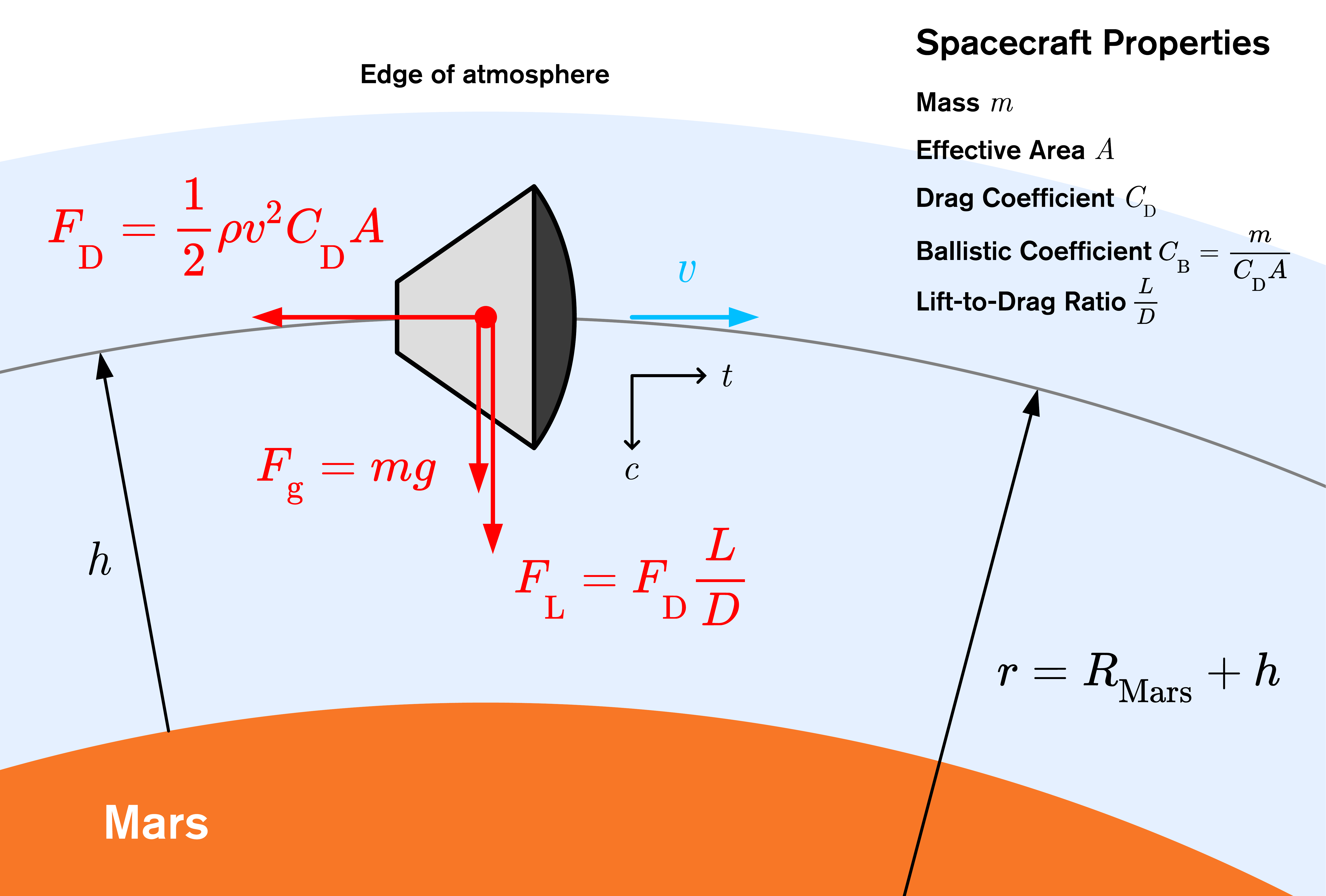}
                    \caption{Idealized aerocapture model}
                    \label{fig:aerocaptureModel}
                \end{figure}
                
                Given the sensitivity of suitable trajectories to several factors (ballistic coefficient, periapsis, lift to drag ratio), an analytical approach was devised to guide trajectory optimization. Lift pointed towards the Martian surface (i.e., negative lift) can be used to force the vehicle's trajectory to remain within the atmosphere longer than would be possible with gravity and drag alone. This effect would be desirable to dissipate the energy of the spacecraft over a longer period, reducing the average thermal load and acceleration felt by the craft. Pushing this effect to its limit leads to solving the following, simpler problem, illustrated in Figure \ref{fig:aerocaptureModel}: For a given spacecraft flying at a particular orbital speed and altitude within the atmosphere, what lift force is necessary to provide enough centripetal acceleration to ensure the craft follows a circular trajectory that matches the curvature of the planet? This model of course is not a faithful representation of atmospheric entry but is useful to narrow the parameter space. Solutions to this problem can be verified through more intensive simulations, rather than testing for all possible configurations.
                
                Considering motion in the centripetal direction, Equation \ref{eq:vcirc} can be derived for the achievable circular velocity $v_\mathrm{circ}$ as a function of the spacecraft's ballistic coefficient $C_\mathrm{B}$ and its lift-to-drag ratio $L/D$. In addition, $v_\mathrm{g-limit}$, the maximum $v_\mathrm{circ}$ for the allowable acceleration felt within the spacecraft, can also be derived by considering the forces transmitted by the spacecraft to its contents (astronaut, rover, etc.).
                
                \begin{align}
                    v_\mathrm{circ}&=\sqrt{\frac{g}{\frac{1}{r}-\frac{\rho}{2C_\mathrm{B}}\frac{L}{D}}}\label{eq:vcirc}\\
                    v_\mathrm{g-limit}&= \sqrt{\frac{2\,a_\mathrm{max}\,C_\mathrm{B}}{\rho\sqrt{1+\left(\frac{L}{D}\right)^2}}}\label{eq:vglimit}
                \end{align}
                Equating both expressions allows us to solve for the atmospheric density required for this maneuver when imposing an acceleration limit (Equation \ref{eq:rholimit}). The required density and associated altitude can be determined by iterating between Equation \ref{eq:rholimit} and the 40NO atmospheric model, and can then be used to solve for the vehicle velocity $v_\mathrm{circ}$. Again, these pseudo-steady-state equations are not meant to be complete solutions to aerocapture---a time-dependent problem---they are used to determine the limits of this maneuver and provide a starting point for simulations.
                
                \begin{figure}[h]
                    \centering
                    \includegraphics[width=\columnwidth]{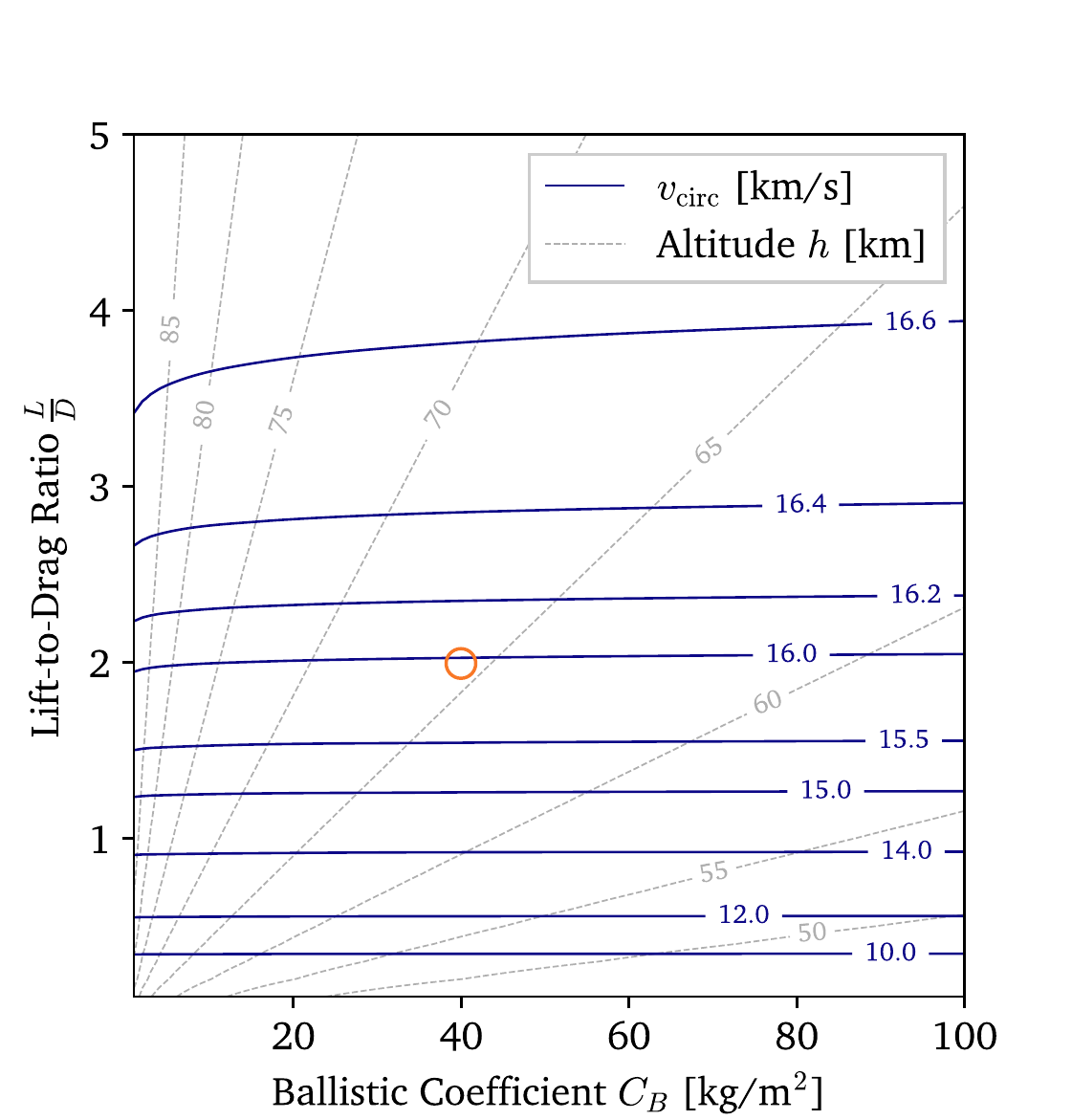}
                    \caption{Idealized aerocapture solutions for an acceleration limit of 8~g. The orange circle identifies the parameters used for trajectory simulation, whose results are plotted in Figure \ref{fig:aerocaptureData}.}
                    \label{fig:aerocapturePlot}
                \end{figure}
                
                \begin{equation}
                    \rho_\mathrm{g-limit} = \frac{2\,a_\mathrm{max}\,C_\mathrm{B}}{r\left(g\sqrt{1+\left(\frac{L}{D}\right)^2}+a_\mathrm{max}\frac{L}{D}\right)}
                    \label{eq:rholimit}
                \end{equation}
                Figure \ref{fig:aerocapturePlot} shows a set of velocity and altitude solutions for an acceleration limit of 8~g, suggesting that while there appear to be viable trajectories below 16~km/s, much greater entry speeds are not feasible with these constraints. This plot was used to estimate and select a subset of parameters in our 2D simulation. With a hyperbolic excess velocity of 15.4~km/s, the incoming spacecraft would enter the Martian atmosphere at approximately 16~km/s, pointing to a required lift-to-drag ratio of two. In practice, a vehicle attempting to perform this would have to continuously adjust its lift-to-drag ratio as it slows down to maintain altitude, then eventually allow itself to rise out of the atmosphere once it has decelerated below the escape velocity. Accordingly, simple $L/D$ modulation laws were implemented in the simulation. Finally, for a 1000-kg conical payload, a ballistic coefficient as low as 40~kg/m$^2$ is thought to be achievable with deployable heat shield \cite{cassell_adept_2018} or ballute technology \cite{rohrschneider_survey_2007}. With these selected properties, trajectories for a range of impact parameters $R_\mathrm{impact}$ were propagated to find a precise aerocapture solution.
            
            \subsubsection{Simulation Results}
\begin{figure}[h]
                    \centering
                    \includegraphics[width=\columnwidth]{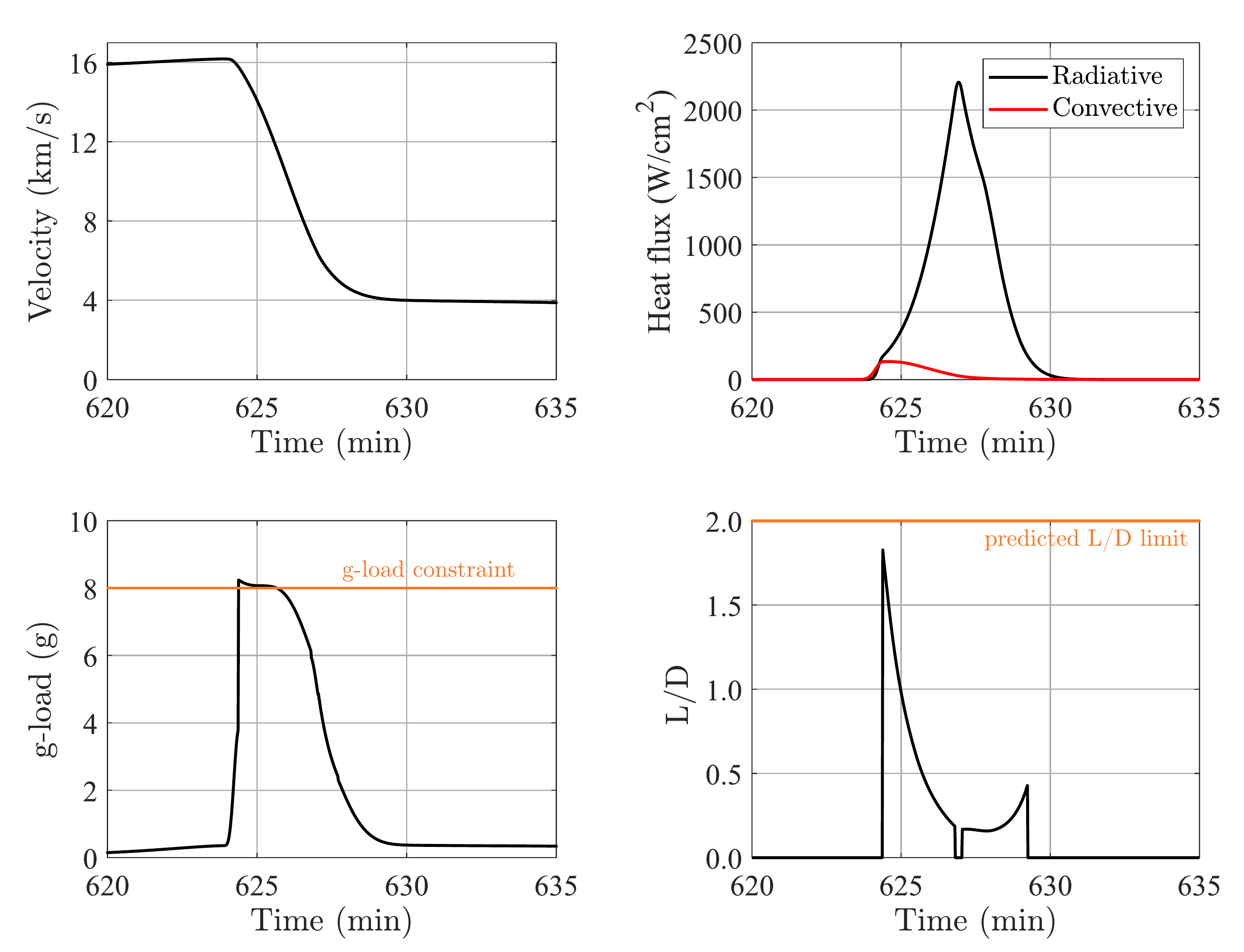}
                    \caption{Aerocapture simulation results. Contact with the atmosphere occurs at approximately 624 minutes of simulation time. Sharp changes in lift-to-drag ratio are due to the use of a simplified control law.}
                    \label{fig:aerocaptureData}
                \end{figure}

                Figure \ref{fig:aerocaptureData} presents plotted data for one of several valid trajectories, showing some agreement with the idealized model (Figure \ref{fig:aerocapturePlot}). The experienced acceleration does not drastically exceed the 8-g limit, and the required lift-to-drag ratio remains below two throughout the maneuver. Although the maximum heat flux, estimated at 2200~W/cm$^2$, is far greater than typical atmospheric entries \cite{wright_mars_2006} and exceeds the capabilities of traditional thermal protection system (TPS) materials \cite{lu_titan_2020}, high-performance TPS are under active development: For example, the Heatshield for Extreme Entry Environment Technology (HEEET), tested at fluxes of up to 3600~W/cm$^2$ \cite{ellerby_heatshield_2019}, appears capable of withstanding the thermal loads of this maneuver.
                
                Although these results are encouraging, the practicality of such a maneuver is still uncertain. Solutions to decrease the ballistic coefficient, sustain intense thermal loads, and modulate a vehicle's lift-to-drag ratio exist independently but may be challenging to integrate into a single vehicle without sacrificing payload capacity. In addition, while the 8-g limit is respected, this load is sustained for several minutes, approaching the limits of human g-force tolerance \cite{creer_centrifuge_1960}, potentially restricting the use of this maneuver to unpiloted systems.

    \section{Spacecraft Architecture}
        The proposed design for the LTPS is based on a 45-day transfer to Mars with a 1-ton payload, although several aspects of the design (e.g., propulsion, reflector) are applicable to any given mission. Key subsystem masses are estimated to provide a lower bound for the LTPS dry mass and its specific mass parameter $\alpha$ (kg/kW). This parameter is a useful metric to compare power-limited propulsion systems such as laser-electric or nuclear-electric propulsion, and it must be properly matched to specific impulse to maximize payload capacity. Ideally, for the missions considered in this study, a 3000-s-$I_\mathrm{sp}$ propulsion system should not exceed an $\alpha$ of 0.005~kg/kW. This requirement is described in detail in \ref{sec:app_IspOpt}.
        
\begin{figure}[t]
            \centering
            \includegraphics[width=\columnwidth]{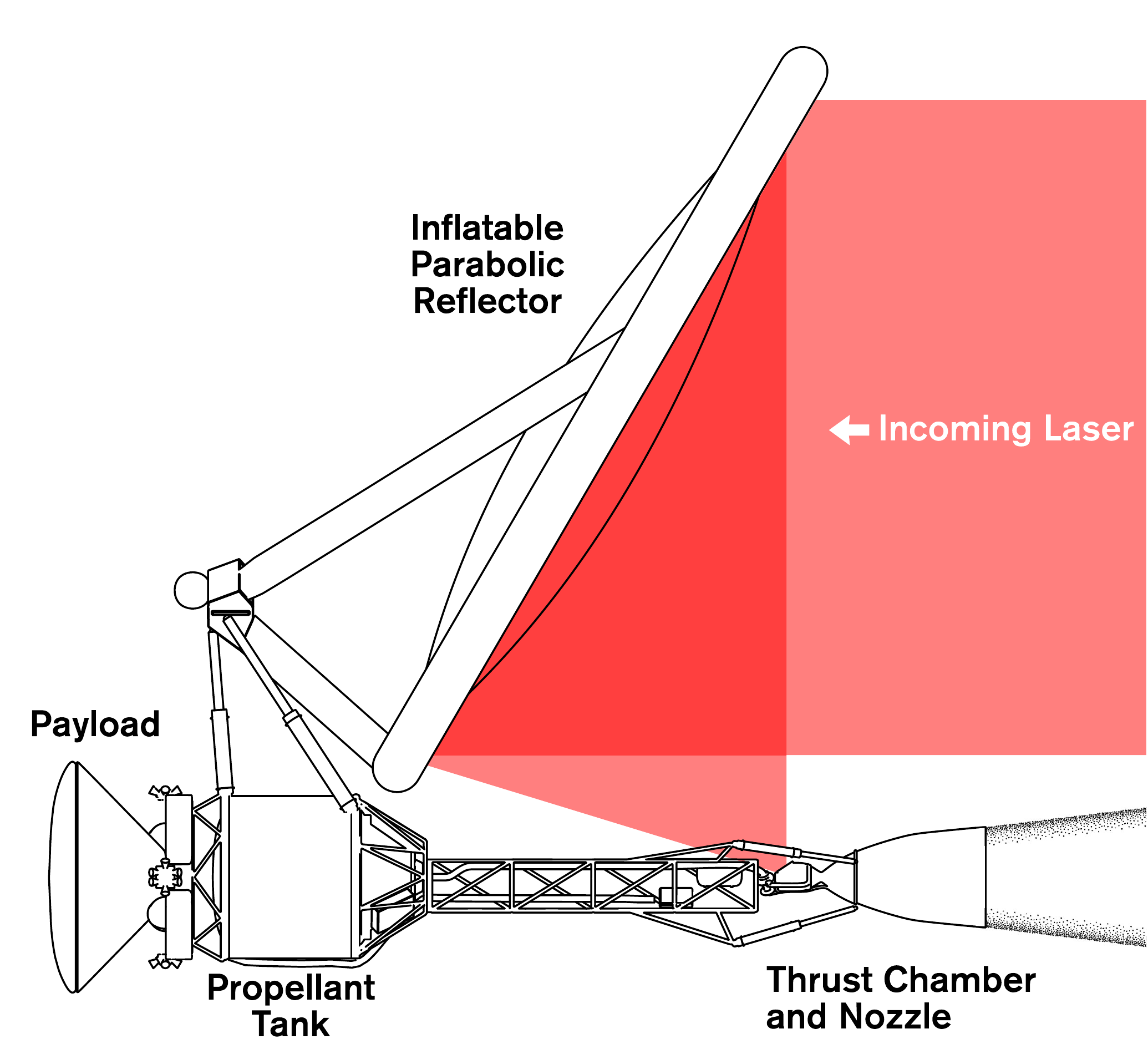}
            \caption{Spacecraft architecture overview}
            \label{fig:overview}
        \end{figure}

        Figure \ref{fig:overview} identifies key subsystems designed for this study and illustrates the basic concept of laser-thermal propulsion: The incoming laser emitted by the array on the ground is collected by a 10-m-wide reflector and focused into the thrust chamber, where the hydrogen propellant is heated then expelled through the nozzle.
                        
        \subsection{Propulsion}
            \subsubsection{Working Principle}
\begin{figure}[t]
                    \centering
                    \includegraphics[width=\columnwidth]{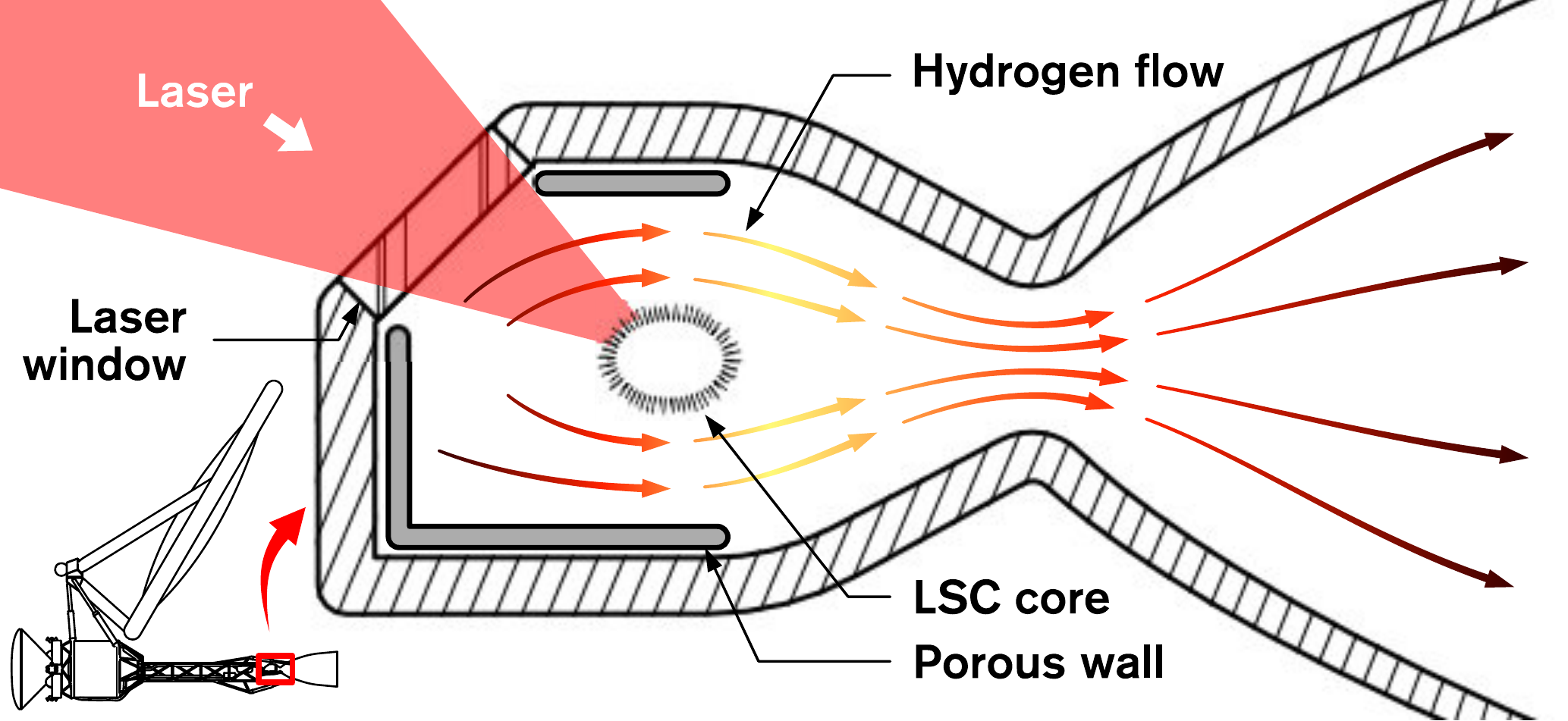}
                    \caption{Conceptual LSC wave thrust chamber schematic}
                    \label{fig:thrustChamber}
                \end{figure}

                The laser thermal system considered for this spacecraft uses a laser to heat a core of hydrogen plasma. As shown in Figure \ref{fig:thrustChamber}, this core is sustained as a Laser-Supported Combustion (LSC) wave and can reach temperatures of 30~000 to 40~000~K in the region of laser deposition. Heat radiating from the core heats up the surrounding flow of gaseous hydrogen, which is then expelled through a high area-ratio nozzle optimized for vacuum operation. This concept is similar in many aspects to nuclear thermal rocket (NTR)---specifically gas-core nuclear rocket (GCNR)---propulsion: Although the power source is different, both systems rely on a plasma core to heat up propellant by radiation, and both would occupy the same performance niche in terms of specific impulse (1000--3000~s). We thus benefited from the extensive literature on NTR and GCNR \cite{rom_nuclear-rocket_1968, poston_heat_1992, kascak_nozzle_1971, kramer_transpiration-cooled_1965}, as LTP faces similar engineering challenges regarding radiation absorption and wall cooling in particular. For a recent review of NTR and GCNR literature, see \cite{warn_roadmap_2021}.
                
                For the purposes of this study, heat transfer and propulsion analyses were performed assuming the stagnation temperature of the flow as equal to its bulk temperature after heat addition. Considering the effects of ionization and dissociation of hydrogen at high temperatures, and assuming complete expansion, it was determined that up to 3000~s of specific impulse could be achievable with a bulk chamber temperature of 10~000~K at 1~atm, requiring a net power input of 90.4~MW for a 0.2~kg/s mass flow rate (see \ref{sec:app_IspCalc} for more details). A previous study by Nored \cite{nored_application_1976} on laser propulsion has found that a 100:1 expansion ratio, for the same chamber temperature at 10~atm, could achieve up to 2500~s of specific impulse. Research on GCNR propulsion suggests that 2500~s is readably achievable at 8300~K \cite{ragsdale_high_1971}. The exact value of specific impulse depends upon factors such as different assumptions regarding nozzle area-ratio or varying chamber pressure. Chamber pressure in particular has a significant effect on specific impulse \cite{watson_nuclear_1994}: Lower pressures allow the hydrogen to dissociate and ionize at lower temperatures, increasing its resulting exhaust velocity. This mission and spacecraft design assumes a 3000-s-$I_\mathrm{sp}$ achieved at 10~000~K, but the effects of lower specific impulses (1000--3000~s) are discussed in \ref{sec:app_IspOpt}. In summary, a reduced $I_\mathrm{sp}$ from 3000 to 2000~s could lead to an 8 to 19\% decrease in payload mass ratio depending on the propulsion system specific mass, should all other aspects of the mission and spacecraft remain unchanged.
            
            \subsubsection{Wall Cooling}
                A key aspect in the design of a laser-thermal thrust chamber is ensuring that most of the radiation emanating from the plasma core is absorbed by the hydrogen, letting as little heat reach the chamber walls as possible. Minimizing this loss is beneficial to maximizing specific impulse and to ensure the integrity of the chamber. GCNR systems tackle an identical issue by seeding the hydrogen propellant with absorbing particles such as carbon or tungsten. This method appears suitable to contain GCNR's 55~000~K uranium plasma cores and provide specific impulses of up to 7000~s \cite{ragsdale_high_1971, ragsdale_gas_1971}. Shoji and Larson \cite{shoji_performance_1976} have applied this concept to laser propulsion and shown that seeding the (otherwise transparent) hydrogen flow with carbon particles can reduce radiation losses to the walls down to 5\% of the incoming laser power (i.e., 4.7~MW). This loss is used as a starting point for the design of cooling solutions for the thrust chamber. While this study assumes the use of propellant-seeding to determine cooling requirements, a specific carbon or tungsten seeding implementation was not considered at this stage of design.
                
                \begin{figure*}[t]
                    \centering
                    \includegraphics[width=\textwidth]{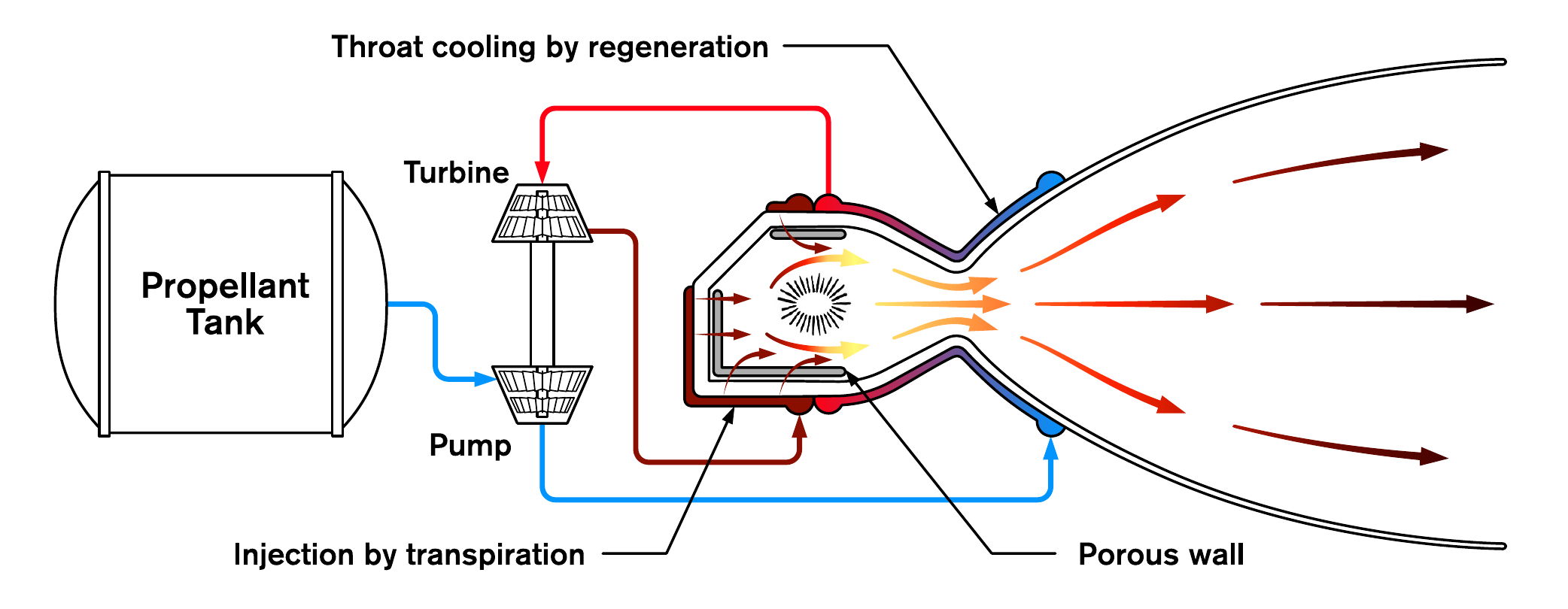}
                    \caption{LTPS feed system diagram}
                    \label{fig:feedsystem}
                \end{figure*}
                
                In order to match the conditions of our LSC model, a nominal chamber geometry of a 1-m-long, 1-m-diameter cylinder made of Inconel X-750 is assumed, with a maximum allowable temperature set to 1000~K to avoid incurring a significant decrease in yield strength \cite{special_metals_corporation_inconelalloy_2004}. Cooling can be ensured in two ways: by running cold propellant in channels in the chamber walls, like many conventional, regeneratively cooled rocket engines, or by forcing the propellant through a porous wall into the chamber, as proposed by some GCNR engine designs \cite{kascak_nozzle_1971}. There is a concern that injecting cold propellant downstream of the thrust chamber would effectively reduce the stagnation temperature of the flow, decreasing the theoretically achievable specific impulse. In either case, ideally, the heat radiated to the walls is no longer lost and is instead used to pre-heat the propellant.
                
                Cooling requirements at the nozzle throat must also be considered, as it is typically subjected to significant thermal loads. The heat transfer coefficient in the nozzle was estimated using the closed-form correlation presented by Bartz \cite{bartz_turbulent_1965}, assuming an approximate conical nozzle profile. The analysis suggests a minimum heat transfer coefficient of around 1400~W/m$^2$-K at the throat to prevent exceeding the thermal limits of Inconel X-750, which would be achievable using conventional regenerative cooling with cryogenic hydrogen. However, due to the low mass flow rate of this design, it may be impractical to use this approach for both the thrust chamber and the nozzle: Should the propellant absorb most of the heat transferred to the walls, its temperature could approach or exceed chamber wall limits, negating its ability to cool the system.
                
                We thus propose the combined use of transpiration and regenerative cooling in an expander cycle, as depicted in Figure \ref{fig:feedsystem}. Pressurized cryogenic hydrogen is fed into cooling channels along the nozzle walls and is then expanded through a turbine to power the pump. The exiting gas is injected into the thrust chamber by transpiration through a porous wall. This hybrid approach has two benefits:
                \begin{itemize}[\textemdash]
                    \item Using regeneration at the nozzle throat instead of transpiration prevents significant reductions in specific impulse since no cold propellant is injected downstream of the LSC core.
                    \item As alluded to at the start of this section, cold hydrogen requires seeding with carbon particles to absorb thermal radiation. The porous wall can serve a similar purpose, absorbing radiation to then transfer it to the cold propellant by convection.
                \end{itemize}
                
                \begin{figure}[t]
                    \centering
                    \begin{subfigure}[b]{\columnwidth}
                        \centering
                        \includegraphics[width=\textwidth]{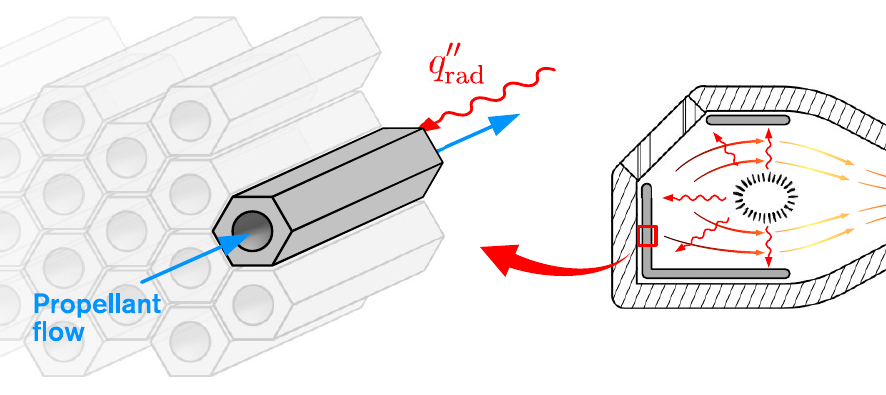}
                        \caption{Simplified model of porous wall structure}
                        \label{fig:ht_porous}
                    \end{subfigure}
                    \begin{subfigure}[b]{\columnwidth}
                        \centering
                        \includegraphics[width=\textwidth]{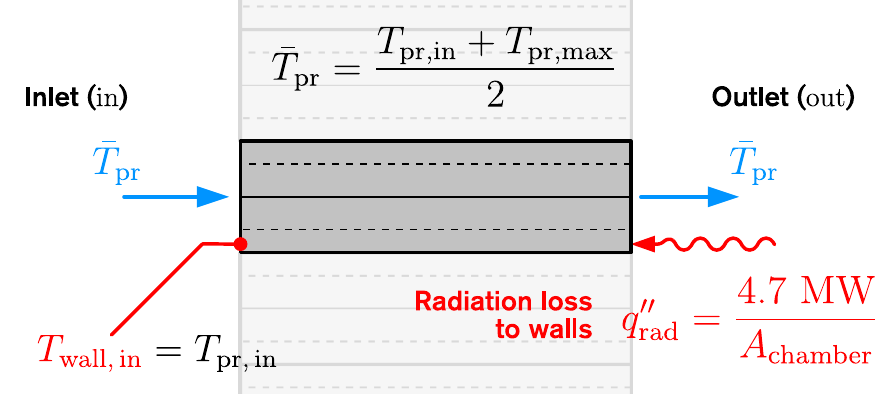}
                        \caption{1D porous wall cooling heat transfer model}
                        \label{fig:ht_model}
                    \end{subfigure}
                    \caption{Porous-wall cooling model}
                    \label{fig:ht}
                \end{figure}
                
                Thermodynamic and heat-transfer calculations were used to determine the performance of this design. By integrating the heat flux computed from Bartz's correlation \cite{bartz_turbulent_1965} (Equation 50), assuming complete coolant-side heat absorption, the maximum outlet temperature of the cooling jacket was calculated to be 522~K, low enough to suggest that it could plausibly maintain the throat temperature below the 1000~K limit.
        
                Isentropic expansion to 1~atm through the turbine lowers the propellant temperature to 382~K, after which it is transpired into the thrust chamber through a porous wall. Previous studies \cite{kascak_nozzle_1971, kramer_transpiration-cooled_1965} have investigated the use of transpiration cooling for GCNR engines operating at greater radiation fluxes, and have found the method suitable, providing confidence in its application to laser-thermal propulsion. For the purpose of mass estimation, a simple porous wall heat transfer model was devised, where the complex pore geometry was approximated as numerous thin tubes stacked next to each other, as shown in Figure \ref{fig:ht_porous}. This one dimensional model could then be solved as a pipe heat transfer problem (Figure \ref{fig:ht_model}), with one pipe boundary condition set to a fixed heat flux of radiated heat from the LSC core. The other boundary's temperature was set to the propellant inlet temperature. One final simplification was the assumption of a constant fluid temperature $\bar{T}_\mathrm{pr}$ at the average between the inlet temperature $T_\text{pr, in}$ and the maximum possible temperature $T_\text{pr, max}$ attainable should the propellant absorb 100\% of the radiated heat. While several parameters affect the resulting inner wall temperature, such as pore geometry and wall thickness, it was found that some configurations prevented the maximum temperature from drastically exceeding the stated 1000~K limit. For instance, a 4-mm-thick, 50\% porous shell was found to reach a maximum temperature of 1044~K. With the addition of an external pressure shell, such a thrust chamber design would weigh 26.8~kg.
                
        \subsection{Feed System}
            The LTPS could operate at low (atmospheric) chamber pressures with a small turbopump, powered by an expander cycle, as described earlier. For the 45-day transfer mission with a $\Delta v$ of 13.95~km/s, assuming an LTPS dry mass of 165~kg (from previous design iterations) and a 1-ton payload, 706~kg of propellant is required, occupying 9.94~m$^3$. Cryogenic hydrogen would be stored in a Kevlar-Epoxy composite tank to minimize mass. Assuming a 2-m-long tank with a 1.3 factor of safety, calculations suggest a dry tank mass of 7.75~kg, or about 1\% of the propellant mass. This lower bound is based on a hoop stress calculation and approximates Kevlar-Epoxy strength assuming a 90° angle between ply fibers, using the expression presented by Bourchak and Harasani \cite{bourchak_analytical_2015}. This hoop stress analysis does not account for a loss of stiffness as the tank empties or vibration loads expected at launch. 
    
            Although cryogenic propellant poses long term storage challenges, this is not relevant in the single-use scenario. A laser-thermal spacecraft could be launched, inserted in its parking orbit, and powered on within a matter of hours, leaving little time for significant propellant leakage. In the re-usable scenario, as shown in Figure \ref{fig:conops}, a refueling procedure would be necessary for each stage-off, performed either by launching the payload with its propellant or with a secondary launch dedicated to refueling. Some form of insulation would nevertheless be necessary, which can be achieved with conventional multi-layer insulation: Assuming 10 layers of aluminized Mylar \cite{institut_fur_arbeitsschutz_der_deutschen_gesetzlichen_unfallversicherung_ifa_polyethylene_nodate}, the insulation increases the tank mass by 8.88~kg.
            
            Low system pressures and mass flow rate suggest that the turbopump mass will be low compared to conventional thrusters. Given that turbomachinery design is out of scope at this stage, an upper bound for its mass was found from existing and proposed turbopump designs operating at a comparable regime. Notably, the estimate generated in \cite{chen_turbopump_2006} indicates that a turbopump with operating conditions on the same order of magnitude would weigh approximately 60~kg, serving as an upper bound for our own mass estimate. Assuming complete absorption by the regeneration system of the heat transferred to the chamber walls, only a small fraction of the heated propellant would need to be expanded through the turbine to power the propellant feed system.
        
        \subsection{Reflector}
            The largest element of the LTPS, its reflector, must be capable of redirecting and focusing the 10~m wide laser spot into the thrust chamber. This task can be performed with an off-axis parabolic mirror, whose optimal focal-length was estimated with the assumption that longer lengths would lead to a linear cost in additional truss elements, while shorter focal lengths increased mirror material costs quadratically. This model yielded a 10$\times$11.6~m paraboloid mirror with a 6-m focal length. The reflector would also feature an active beacon, providing the laser array with a co-operative target to facilitate tracking.
            
            \begin{figure}[t]
                \centering
                \includegraphics[width=\columnwidth]{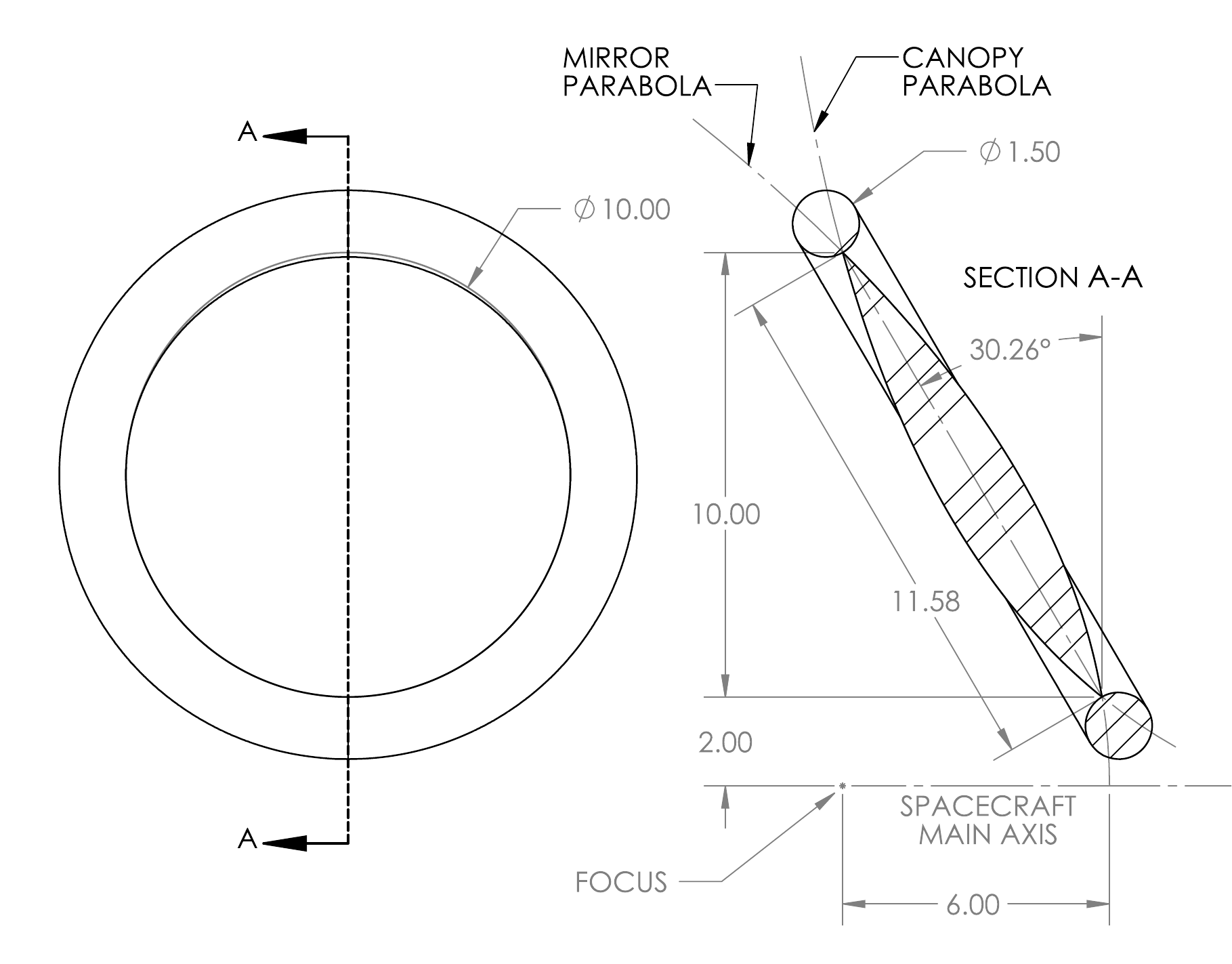}
                \caption{Reflector design and nominal dimensions [m]}
                \label{fig:reflector}
            \end{figure}
    
            Of course, a rigid mirror of this size would be too heavy and cumbersome to deploy or fit inside a payload fairing. Our design considered the use of a lightweight, inflatable reflector instead. Inflatable space structures are under active development due to their low mass, compact launch package and simple deployment. Several inflatable reflectors have already been constructed \cite{engberg_modal_2000}, flown, and tested in orbit \cite{freeland_large_1997}, for telecommunications and solar thermal propulsion applications \cite{jacob_inflatable_2020}.
            
            The reflector follows a similar design to off-axis concentrators studied for solar thermal spacecraft \cite{grossman_inflatable_1990, grossman_analysis_1991}, with two identical paraboloid membranes forming the main lenticular body. The mirror-side membrane would be coated in several dielectric layers (TiO$_2$/SiO$_2$ or Ta$_2$O$_5$/SiO$_2$) to achieve 99.5\% to 99.8\% reflectivity, respectively \cite{watkins_optical_1993}. The other membrane, the canopy, only exists to form an enclosed pressurized space, and must be transparent to 1.06-µm radiation both to reduce losses and remain within its operational temperature range. Table \ref{tab:reflectorTemp} lists the absorptivity $\alpha$, the emissivity $\epsilon$, and the approximate equilibrium temperature $T_\text{eq}$ of potential canopy and mirror materials\footnote{PI-6FDA absorptivity and emissivity taken from \cite{xiao_high_2017}, where absorptivity was scaled for the reflector's membrane thickness. TiO$_2$/SiO$_2$ absorptivity was computed from \cite{watkins_optical_1993} assuming no transmission and its emissivity value was taken from \cite{yang_preparation_2011}.} under 90~MW of laser power, for a thickness of 0.0254~mm. Note that the canopy is exposed to both the incoming laser flux and its reflection from the mirror, doubling the effective incident power to 180~MW. Fluorinated polyimide films (PI-6FDA) were selected as a transparent canopy material for their low absorptivity, allowing them to tolerate the intense laser flux while remaining well below their high glass-transition temperature of 321°C \cite{xiao_high_2017}. This heat-resistance also makes this polyimide suitable as a substrate for the dielectric mirror coatings. The lenticular mirror would be supported by inflatable struts and tensioning torus made of the same material. Modest internal pressures from 2 to 150~Pa are sufficient to maintain the shape of the mirror in the vacuum of space, and are preferable to minimize the flow rate of potential leaks.
        
            \begin{table}[t]
                \centering
                \caption{Reflector materials and their equilibrium temperature}
                \label{tab:reflectorTemp}
                \begin{tabular}{@{}llrrr@{}}
                    \toprule
                    Material        & Function & $\alpha$ & $\epsilon$ & $T_\text{eq}$ {[}°C{]} \\ \midrule
                    PI-6FDA & Canopy   & 0.0017   & 0.600      & 195                    \\
                    TiO$_2$/SiO$_2$ & Mirror   & 0.0050   & 0.685      & 230                    \\ \bottomrule
                \end{tabular}
            \end{table}

            A concern with the use of an inflatable reflector is the impact of wrinkles and other deformations on optical performance. Optical performance tests of prototype inflatable solar concentrators reported slope errors of 3~milliradians rms \cite{grossman_inflatable_1990}, which would translate to a focal spot radius of approximately 3~cm for the reflector envisioned in this study, which is satisfactory. Another concern, particularly for propulsion applications, is the ability of the inflatable structure to retain its shape and position while accelerating. For the 45-day Mars transfer, spacecraft acceleration is not expected to exceed 1~g. The inflatable reflector must be able to maintain its focus close to the center of the thrust chamber under this acceleration. Inflatable beam deflection equations \cite{suhey_numerical_2005} were used to provide an order of magnitude estimate of the deflection of the reflector at its tip under 10 m/s$^2$ of acceleration. This deflection was found to be 4~mm, negligible given the overall size of the reflector, and results in a proportionally small change in focal point. A more complete study of reflector deformation would be needed to understand its effects on the stability of its focal point, and whether actuators could provide sufficient control to compensate. \rev{Active control mechanisms for the reflector would play a critical role in a fully operational LTPS. In addition to stabilizing the reflector, orientation mechanisms and secondary optics would be necessary to allow the spacecraft to decouple its thrust direction from the incoming beam direction. The design presented in Figures \ref{fig:ltpRender} and \ref{fig:overview} clearly presents the working principles of laser-thermal propulsion, but is limited to thrusting along the incoming laser's direction.}
            
            Figure \ref{fig:reflector} indicates nominal reflector dimensions, allowing for a mass estimate based on its computed surface area and volume. Required membrane thicknesses were selected based on  the 2 and 150~Pa pressures in the lenticular body and torus, respectively, and commercially available film thicknesses. Inflated with helium, the reflector would add 39.6~kg to the LTPS mass.

        \subsection{Mass and Specific Mass}
            \begin{table}[b]
                \centering
                \caption{Dry LTPS mass summary. This omits several spacecraft subsystems necessary for a complete design, but is used to establish a lower bound for the specific mass of the LTPS.}
                \label{tab:masses}
                \begin{tabular}{@{}lr@{}}
                    \toprule
                    Component                    & Mass {[}kg{]} \\ \midrule
                    Thrust chamber               & 26.80         \\
                    Propellant tank              & 7.75          \\
                    Tank insulation              & 8.88          \\
                    Turbopump assembly           & 60.00         \\
                    Reflector membrane           & 39.59         \\
                    Reflector pressurization gas & 0.03          \\ \midrule
                    Total                        & 143.05        \\ \bottomrule   
                \end{tabular}
            \end{table}
            Table \ref{tab:masses} provides a mass summary of our design, totaling at 143~kg, which is of course an absolute lower bound for the overall propulsion system mass. Several components are missing from this calculation: such as regeneration jackets, truss elements, piping, and more. Nevertheless, this allows for an initial calculation of the LTPS specific mass parameter $\alpha$, keeping in mind that a value of 0.005~kg/kW should not be exceeded to maximize payload capacity as shown in \ref{sec:app_IspOpt}.
    
            For a gross power input of 90~MW and a 143-kg dry system mass, we find an $\alpha$ of 0.0016~kg/kW, far lower than needed. An ample mass margin is thus left for neglected subsystems mentioned earlier. Indeed, the LTPS mass could increase by an order of magnitude and still remain competitive. In fact, any $\alpha$ below 1~kg/kW could enable rapid-transit missions within the solar system \cite{pelaccio_examination_2002}.

    \section{Alternative Missions}
        \label{sec:altMissions}
        Although preliminary calculations show promise in the design and performance of an LTP transfer stage, simulations suggest that using this architecture to reach Mars within 45 days requires---at a minimum---a very delicate aerocapture maneuver if no laser array is available to effect the deceleration maneuver. Should this fail, while the trajectory discussed in Section \ref{sec:mars1} could be tuned to provide a free return (orbital period of 3.96 years), the duration of such a return makes it impractical. Nevertheless, the high specific impulse of LTP still makes this system attractive to increase the payload capacity of a mission using more conventional Hohmann transfers.
        
        Table \ref{tab:altMissions} features propellant ($m_\mathrm{pr}$) and payload ($m_\mathrm{pl}$) mass data for three missions, powered by chemical or LT propulsion. \textbf{Mars~1} represents the 45-day mission described in detail in this paper. \textbf{Mars~2a} considers a similar mission for a piloted spacecraft, including adequate life support systems for the outgoing trip, estimated to weigh at least 40 tons (based on Orion capsule and European Service Module wet mass). Finally, \textbf{Mars~2b} attempts to carry as much payload as possible with the propellant available in a single-engine Centaur, utilizing a typical Hohmann transfer. Mass data was calculated for three propulsion systems: the LTPS as described in this study, a heavier LTPS with $\alpha=0.005$~kg/kW, and a single-engine Centaur upper stage \cite{united_launch_alliance_atlas_2010} for comparison.
        
        \begin{table*}[t]
            \centering
            \caption{Alternative Mars transfers with single use or re-usable stages. $m_\text{pl}$: payload mass, $\Delta{v}$: mission delta-v from medium Earth orbit, $\alpha$: specific mass, $m_\text{ps}$: propulsion system mass, $I_\text{sp}$: specific impulse, $m_\text{pr}$: propellant mass.}
            \label{tab:altMissions}
            \begin{tabular}{@{}lrrrlrrrrrr@{}}
            \toprule
            Mission                  & \begin{tabular}[c]{@{}r@{}}Time\\ {[}days{]}\end{tabular} & $m_\text{pl}$ {[}kg{]}  & \begin{tabular}[c]{@{}r@{}}$\Delta v$\\ {[}km/s{]}\end{tabular} & \begin{tabular}[c]{@{}l@{}}Propulsion\\ System\end{tabular} & \begin{tabular}[c]{@{}r@{}}Power\\ {[}MW{]}\end{tabular} & \begin{tabular}[c]{@{}r@{}}$\alpha$\\ {[}kg/kW{]}\end{tabular} & $m_\text{ps}$ {[}kg{]} & $I_\text{sp}$ {[}s{]} & \begin{tabular}[c]{@{}r@{}}$m_\text{pr}$ {[}kg{]}\\ (Single)\end{tabular} & \begin{tabular}[c]{@{}r@{}}$m_\text{pr}$ {[}kg{]}\\ (Reuse)\end{tabular} \\ \midrule
            \multirow{3}{*}{Mars 1}  & \multirow{3}{*}{45}                                       & \multirow{3}{*}{1 000}  & \multirow{3}{*}{13.95}                                          & LTPS                                                        & 100                                                      & 0.0016                                                         & 160                    & 3 000                 & 700                                                                       & 860                                                                      \\
                                     &                                                           &                         &                                                                 & LTPS                                                        & 100                                                      & 0.0050                                                         & 500                    & 3 000                 & 910                                                                       & 1 400                                                                    \\
                                     &                                                           &                         &                                                                 & Centaur                                                     & N/A                                                         &  N/A                                                              & 2 247                  & 451                   & 73 000                                                                    & 1 300 000                                                                \\ \midrule
            \multirow{3}{*}{Mars 2a} & \multirow{3}{*}{45}                                       & \multirow{3}{*}{40 000} & \multirow{3}{*}{13.95}                                          & LTPS                                                        & 4 000                                                    & 0.0016                                                         & 6 400                  & 3 000                  & 28 000                                                                    & 34 000                                                                   \\
                                     &                                                           &                         &                                                                 & LTPS                                                        & 4 000                                                    & 0.0050                                                         & 20 000                 & 3 000                  & 36 000                                                                    & 56 000                                                                   \\
                                     &                                                           &                         &                                                                 & Centaur                                                     &  N/A                                                        &  N/A                                                              & 2 247                  & 451                   & 950 000                                                                   & 2 100 000                                                                \\ \midrule
            \multirow{3}{*}{Mars 2b} & \multirow{3}{*}{180}                                      & 130 000                 & \multirow{3}{*}{4.18}                                           & LTPS                                                        & 3 000                                                    & 0.0016                                                         & 4 800                  & 3 000                 & 20 830                                                   & 22 000                                                                   \\
                                     &                                                           & 120 000                 &                                                                 & LTPS                                                        & 3 000                                                    & 0.0050                                                         & 15 000                 & 3 000                 &                     20 830                                                      & 23 000                                                                   \\
                                     &                                                           & 11 000                  &                                                                 & Centaur                                                     &  N/A                                                        &  N/A                                                              & 2 247                  & 451                   &                   20 830                                                        & 30 000                                                                   \\ \bottomrule
            \end{tabular}
            \end{table*}
        
        The use of LTP could lead to a 10-fold increase in payload capacity compared to chemical thrusters for the same propellant mass. These capabilities make this architecture especially attractive in the context of long-term Martian settlements, where large quantities of specialized equipment, habitats, and consumables sent from Earth will be needed to support a colony. Should such a settlement construct its own directed-energy system, laser-thermal transfer stages could become the workhorse of an interplanetary economy, providing means for both fast transit and large cargo shipments between Earth and Mars.

        In addition, accessibility to more distant targets beyond the asteroid belt and to the edge of the solar system could be greatly improved with the use of LTP. Exploration missions to gas giants could be performed via a direct Hohmann transfer, reducing mission time and increasing launch window frequency. Furthermore, the LTPS as described here could be used as a solar-thermal spacecraft with little modification. A solar Oberth maneuver could be performed with the same spacecraft, potentially enabling the flyby of interstellar objects such as 1I/’Oumuamua \cite{hein_project_2019, hein_interstellar_2021}. Finally, this architecture is also suitable for interstellar precursor missions, such as placing a spacecraft at the solar gravitational focus. Beyond this point, 550~AU away, light from distant star systems is focused by the Sun's gravity, potentially enabling megapixel resolution imaging of exoplanets \cite{turyshev_direct_2020}. Such interstellar precursor missions require $\Delta v$'s of 30--50 km/s, which is within the capability of laser-thermal propulsion featuring $I_\mathrm{sp} \approx 3000$~s. These missions—which are typically flybys in nature and do not require rendezvous with a target—are better suited to the laser-thermal architecture explored in this study than payload delivery to Mars, since flybys only require significant $\Delta v$ at departure from Earth.

    \section{Discussion and Further Work}
        Enabled by shorter laser wavelength and the ability to operate as a phased array of unprecedented optical dimensions, laser-thermal propulsion can now be extended two orders of magnitude deeper into cislunar space than previously considered in the 1970s and 1980s. A second advantage that this proposed architecture capitalizes upon is the laser fluxes that are permissible upon the inflatable reflector, which exceed by two orders of magnitude the flux limitations on laser-electric propulsion with no active cooling \cite{sheerin_fast_2021,brophy_directed-energy_2019}. These high fluxes allow laser-thermal propulsion to “burn hard” early in the mission, while the spacecraft is still within the focal length of the laser, enabling high $\Delta v$ missions with 10-m-scale lasers. A further benefit of the ability to perform high thrust burns in near-Earth space is the propulsion stage (which includes the hydrogen tank, heating chamber, nozzle, and reflector) can be immediately brought back to Earth, where it can be rapidly refueled and reused during the same launch window. While incurring a modest loss in payload  capacity, the re-usability of the LTPS offers significant advantages over allowing the hardware to proceed to Mars, where it would have little utility.

        By borrowing and building upon concepts developed for solar-thermal and gas-core nuclear propulsion, our proposed design for an LTPS appears plausible and promises an unprecedented mass-to-power ratio. In fact, the values of $\alpha$ found in this study are so low that they no longer influence the mission design; even if $\alpha$ values must increase by an order of magnitude as the design of the propulsion system is further refined (see Table \ref{tab:altMissions}), the implications for the ability to meet the mission objectives with a significant payload fraction are negligible. In fact, a specific impulse of 3000~s is optimal for a propulsion system three to four times heavier ($\alpha \approx 0.005$ kg/kW), as detailed in \ref{sec:app_IspOpt}.

        The stability and radiative heating properties of the laser-supported plasma should be further studied. Plasma instability in particular is a complex phenomenon and is critical to this propulsion system, as the inflatable reflector and tracking inaccuracies could be sources of instability. Numerical simulation and experimental work is underway in our research group to study the radiative properties of hydrogen plasma, the implementation of particle-seeding, and its effects on heat-transfer. Small-scale experiments on laser-supported plasma have been performed for propulsion applications \cite{inoue_oscillation_2004} and could be expanded to study plasma stability and its response to disturbances stemming from imperfect optics. \rev{Such experiments, coupled with extensive simulations, would pave the way to small-scale thruster prototypes operated by individual lasers within a laboratory setting. Given the existing work on laser-supported plasma, we believe the realization of such prototypes may be feasible within the next decade.}
        
        Should continued research efforts lead to the development of functional prototype LTP thrusters, smaller-scale test missions could be envisioned within low to medium Earth orbit, evaluating the performance of laser-tracking and the inflatable reflector. \rev{Orbital Transfer Vehicle (OTV) applications, already considered for solar-thermal propulsion \cite{engberg_modal_2000}, could also be served by small-scale laser-thermal thrusters. Such applications provide a useful de-scoping option that can still compete against chemical propulsion in terms of propellant efficiency.} The scalability of LTPS spacecraft should be further studied, as the focusing limits (and therefore the maximum thrust duration) of the laser array will depend on the reflector size. Several other inter-dependent mission parameters such as laser power, parking orbit, or propellant flow rates can be adjusted to compensate for focusing limits, motivating the need for an in-depth analysis.
        
        The trade space between laser-electric and laser-thermal propulsion (\ref{sec:app_IspOpt}) should be further explored: Laser electric has the significant advantages of a greater specific impulse and a modular architecture—permitting all components to be tested and validated on a benchtop—but at the expense of a limited laser flux that the photovoltaics that are used to convert laser power to electricity can tolerate due to thermal constraints. The study of Sheerin et al. used a value of 10 kW/m$^2$ (i.e., about seven suns), which was deemed feasible without the use of active cooling of the photovoltaics. As a consequence, the propulsive maneuver for laser-electric propulsion missions typically requires days to weeks, which for continuous power delivery would necessitate multiple laser sites on Earth or construction of a laser array in space. The laser-electric missions considered in \cite{sheerin_fast_2021} also necessitated larger arrays (750 to 1000 m in effective diameter) in comparison to the present study’s 10-m array for the same class of payload. Our preliminary conclusion is that, as larger laser arrays become available, laser-electric offers the greatest benefits, but for early application of phased-array lasers with 10-m-scale laser arrays, laser-thermal may offer a greater potential to realize missions with significant payloads.
        
        This study has also shown that LTP greatly benefits from advances made in gas-core nuclear propulsion and breakthroughs in directed-energy concepts. These links to alternate propulsion systems and the application-agnostic nature of laser-arrays should motivate the study of possible synergies between LTP, LEP, and perhaps even NTR propulsion systems. Hybrid directed-energy systems could perhaps allow a spacecraft to benefit from the best of both worlds: using a high-thrust, short duration laser-thermal maneuver to quickly escape Earth's sphere of influence, then discarding the LTP hardware to rely on lower-power but more efficient LEP for the rest of the mission. Common plasma stability and heat-transfer challenges are faced by both LTP and GCNR propulsion, and the development of solutions to these issues would benefit both architectures. \rev{The renewed interest in nuclear-thermal propulsion \cite{skelly_nasa_2021, skelly_nuclear_2021} could be an opportunity to resolve these shared developmental issues in parallel. Cooling technologies designed for GCNR propulsion can be re-purposed for LTP, and work on laser-supported plasma stability can benefit GCNR development without requiring fissile material for experiments.} There are also some commonalities for both laser-thermal and laser-electric architectures considered in the present study: Since a laser of sufficient power for directed-energy propulsion would not yet be available on Mars, any rapid LTP or LEP Mars mission performed with a sub-km laser array would have to resort to aerocapture in order to land or orbit the planet. The present study suggests that aerocapture with an approach hyperbolic excess velocity of 15 km/s is feasible with current thermal protection materials and vehicle designs within the state of the art. Several of the LEP flyby missions presented in \cite{sheerin_fast_2021} could thus be turned into orbit transfer missions. Aerocapture maneuvers could be considered for any target with a sufficiently dense atmosphere. More detailed aerocapture analyses, such as ones done for Venus \cite{lockwood_systems_2006}, Mars \cite{wright_mars_2006}, Titan \cite{lockwood_titan_2003}, and Neptune \cite{lockwood_aerocapture_2006}, could be revisited within the context of directed-energy missions to greatly expand the range of missions achievable with single, sub-km directed-energy arrays.

    \section{Conclusion}
        The implications of the emergence of phased-array lasers of 10-m-scale and 100-MW power for the design of a high $\Delta v$ mission have been examined, and the results of this study suggest the potential for a disruption in comparison to conventional chemical and solar-electric propulsion. The high specific impulse achieved with directed energy allows laser-thermal propulsion to perform interplanetary missions with less propellant than chemical systems and in shorter thrust durations than solar-electric propulsion.

        The preliminary design of critical subsystems necessary for such a spacecraft has not found fundamental technological roadblocks to realize this propulsion system. Furthermore, the mass-to-power ratios ($\alpha$) values that may be achieved via laser-thermal propulsion (0.001--0.010 kg/kW) are unparalleled, far below even those cited for advanced nuclear propulsion technologies, due to the fact that the power source remains on Earth and the delivered flux can be processed by a low-mass inflatable reflector.
        
        We reiterate that the Mars-in-45-days goal is only used as a convenient metric for propulsion architectures. This requirement is motivated by the ability to deliver astronauts to the lower radiation environment on the surface of Mars while absolutely minimizing exposure to GCRs and potential CME events in route. Cargo delivery missions were explored as well as part of this study, showing a potential ten-fold increase in payload capacity compared to the Common Centaur cryogenic upper stage. Other missions of interest previously considered unfeasible with conventional chemical or solar-electric propulsion may be realizable via this architecture, such as: rapid missions to the outer ice giant planets, into the interstellar medium, intercepting interstellar objects passing through the solar system, and to the solar gravitational focus.
    
    \section{Acknowledgements}
        The authors would like to acknowledge the valuable assistance of Mathias Larrouturou, Lynn Cherif, Rahul Atmanathan, Samuel Smocot, and Alp Tanriover in developing this design study. We would also like to thank Philip Lubin, Carl Knowlen, Adam Bruckner, Mélanie Tétreault-Friend, and all participants to this study's Preliminary Design Review held in August 2020 for their input on this project. Our thermodynamic analyses were made possible by the \emph{CoolProp} Open-Source Thermophysical Property Library \cite{bell_pure_2014} and NASA Chemical Equilibrium with Applications (CEA) data \cite{mcbride_nasa_2002}. Porkchop plots (Figure \ref{fig:porkchop}) were generated using resources provided in \cite{yaylali_david_2019}. This work was supported by the Natural Sciences and Engineering Research Council of Canada (NSERC) Discovery Grant “Dynamic Materials Testing for Ultrahigh-Speed Spaceflight” and the McGill Summer Undergraduate Research in Engineering program.

    \appendix
        \section{Specific Impulse Optimization}
            \setcounter{figure}{0}
            \label{sec:app_IspOpt}
            
            Much like electric propulsion, a laser-thermal thruster is power-limited by its energy source, such that its payload mass fraction $m_\mathrm{pl}/m_0$ can be maximized as shown by Stuhlinger \cite{stuhlinger_electric_1967}, based on the propulsive maneuver duration $t_\mathrm{m}$ and mission $\Delta v$. The relevant equation is reproduced here as Equation \ref{eq:stuhlingerOpt}:
            \begin{equation}
                \frac{m_\mathrm{pl}}{m_0} = \mathrm{e}^{- \frac{\Delta v^*}{c^*}}-{c^*}^2\Big(1-\mathrm{e}^{- \frac{\Delta v^*}{c^*}}\Big)
                \label{eq:stuhlingerOpt}
            \end{equation}
            Where normalized delta-v $\Delta v^*$ and normalized exhaust velocity $c^*$ are expressed based on the characteristic velocity $v_\mathrm{ch}$, defined by Equation \ref{eq:vch} in terms of maneuver duration $t_\mathrm{m}$ and the power and propulsion system (PPS) specific mass $\alpha$ [kg/W]. It is assumed that the PPS mass is proportional to the resulting jet power.
            \begin{equation}
                v_\mathrm{ch} = \sqrt{\frac{2\,\eta\,t_\mathrm{m}}{\alpha}}
                \label{eq:vch}
            \end{equation}
            \begin{equation}\Delta v^* = \frac{\Delta v}{v_\mathrm{ch}}\end{equation}
            \begin{equation}c^* = \frac{c}{v_\mathrm{ch}} = \frac{I_\mathrm{sp}\,g_0}{v_\mathrm{ch}}\end{equation}
            These equations quantify the trade-off between specific impulse and thrust, and imply that a greater $I_\mathrm{sp}$ is not always desirable.
            
\begin{figure}[t]
                \centering
                \includegraphics[width=\columnwidth]{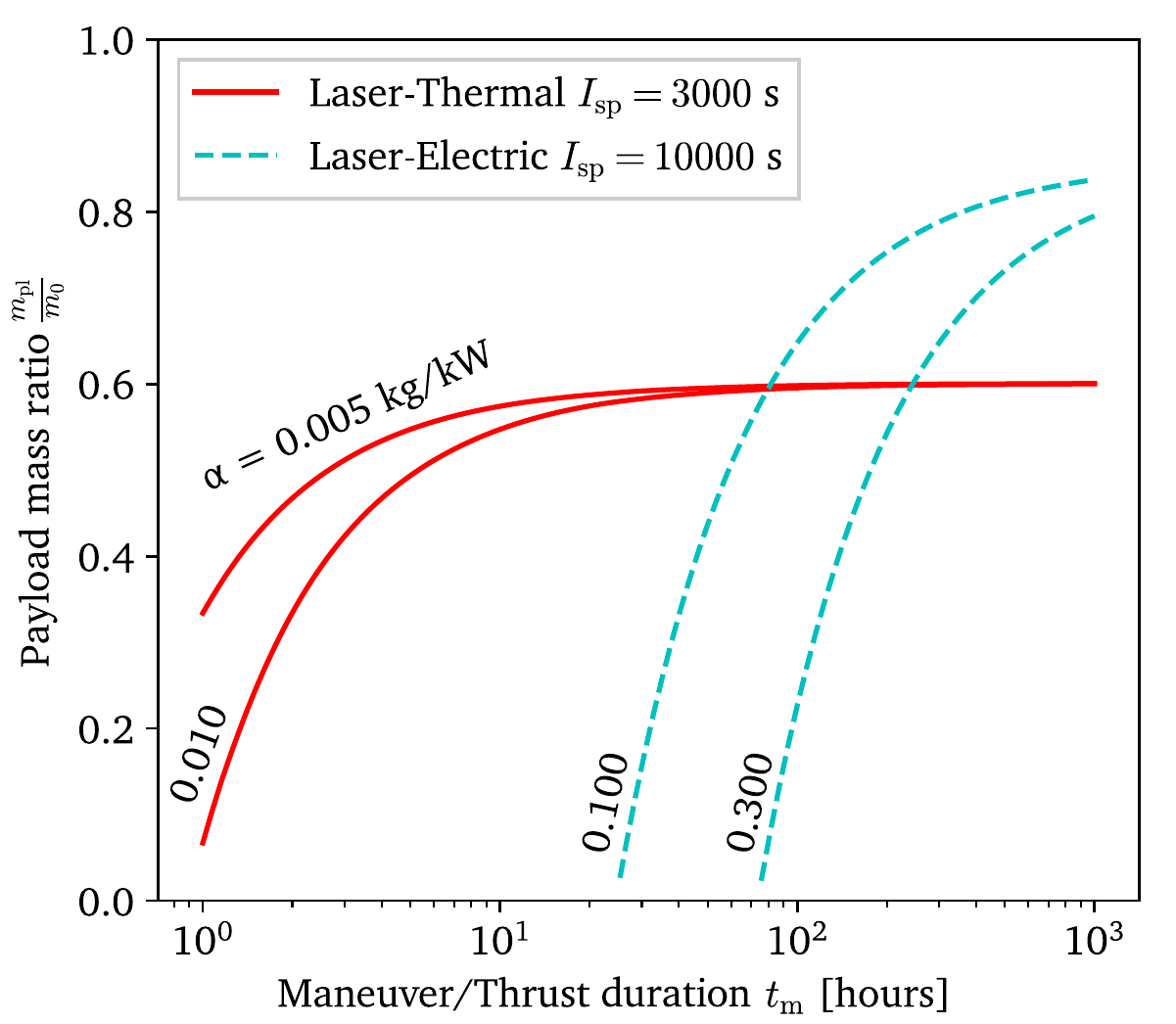}
                \caption{Comparison of payload capacity vs. thrust duration for different propulsion systems for $\Delta v=15$~km/s. Expected $\alpha$ for laser-electric systems taken from \cite{sheerin_fast_2021}}.
                \label{fig:app_stuhlComp}
            \end{figure}

            For example, different propulsion systems, characterized by their specific mass $\alpha$, can be compared for a 45-day Mars transit in terms of payload mass fraction and thrust duration. For a mission $\Delta v$ of 15~km/s and an efficiency $\eta$ of 90\%, the payload mass fraction of both laser-thermal and laser-electric propulsion systems are plotted in Figure \ref{fig:app_stuhlComp}. Fixed specific impulses were assumed for this comparison, assuming 10~000~s for laser-electric systems as per \cite{sheerin_fast_2021}.
            
            \begin{figure}[h]
                \centering
                \includegraphics[width=\columnwidth]{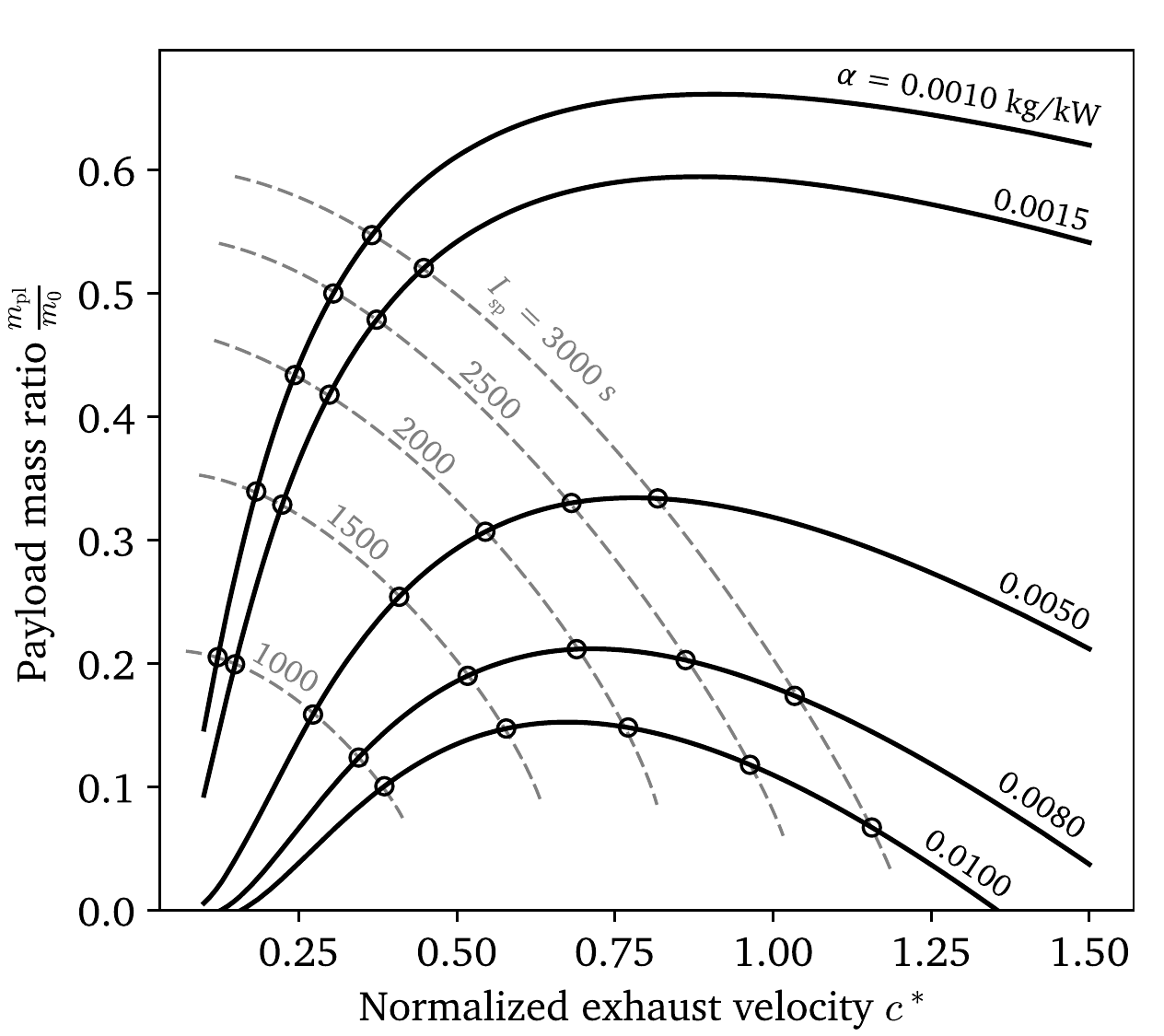}
                \caption{Payload mass ratio dependency on normalized specific impulse for $\Delta v=15$~km/s. Constant $I_\mathrm{sp}$ lines (dashed) reveal the sensitivity of the payload mass ratio to the specific impulse for a given $\alpha$}
                \label{fig:app_stuhlOpt}
            \end{figure}
            
            The effect of the specific mass on payload capacity is made clear in Figure \ref{fig:app_stuhlComp}. Where laser-thermal-propulsion systems may be able to launch significant payloads with one to two hours of thrust, heavier systems such as laser-electric propulsion would require 10 to 100 hours to have a positive payload mass ratio. However, when such maneuver durations are acceptable and feasible, laser-electric systems quickly overtake LTP owing to their typically greater specific impulse.
            
            Equations \ref{eq:stuhlingerOpt} and \ref{eq:vch} can also be used to optimize specific impulse for a given PPS specific mass. For time constrained missions, the payload mass ratio reaches a maximum for an exhaust velocity close to the characteristic velocity. Beyond that, increased exhaust velocity prevents the propulsion system from providing enough thrust to accelerate the spacecraft within the time constraint, reducing the payload mass fraction of the system.
            
            In the context of developing a laser infrastructure to eventually power interstellar missions, finding applications for directed-energy propulsion feasible with a single, small array is desirable, motivating a 1-hour constraint on the propulsive maneuver. The intended missions for laser-thermal propulsion, such as rapid transit to Mars or direct transit to the edge of the solar system, require $\Delta v$ ranging from 10 to 15~km/s. These parameters allow us to generate Figure \ref{fig:app_stuhlOpt}, where the payload mass ratio is plotted as a function of normalized exhaust velocity and specific mass. A range of specific impulse values is overlaid to reveal the sensitivity of the payload mass ratio to $I_\mathrm{sp}$. The design presented in this study, with an $I_\mathrm{sp}$ of 3000~s and a specific mass close to 0.0015~kg/kW would achieve a payload capacity of about 50\%. However, an even greater exhaust velocity would be desirable at this specific mass to maximize the performance of this system, and it is unlikely that a fully realized LTPS would be as light as estimated here.
            
            Fortunately, it appears as though a 3000-s specific impulse is close to the optimal exhaust velocity for a specific mass of 0.005~kg/kW. Thus, an LTP system optimized for rapid transit missions in the solar system with a 1-hour thrust duration should have a specific mass of 0.005~kg/kW, hence its inclusion in Table \ref{tab:altMissions} as an optimal LTPS. Should a practical LTPS have an even greater $\alpha$, such as 0.01~kg/kW, the specific impulse of the system could be easily reduced by increasing propellant mass flow rate for the same input power, in order to maximize the payload mass ratio.
            
            An additional benefit of a greater specific mass is a lower sensitivity to changes in specific impulse. For instance, a reduction in $I_\mathrm{sp}$ from 3000 to 2000~s results in a 19\% decrease in payload mass ratio for a 0.0015-kg/kW-$\alpha$, compared to an 8\% decrease for a 0.0050-kg/kW-$\alpha$. While these performance losses are not negligible, the payload capacity enabled by LTP remains considerable compared to chemical systems.
            
        \section{Specific Impulse of Laser-Heated Hydrogen}
            \label{sec:app_IspCalc}
            The specific impulse of laser-heated hydrogen propellant was determined by converting the enthalpy of the propellant in the heating chamber into kinetic energy, i.e., velocity, of the exhaust. The performance was bounded by considering two cases: An ideal case with chemical equilibrium maintained throughout the nozzle expansion and a worst-case scenario where the chemical composition throughout the nozzle is kept ``frozen'', i.e., no recombination occurs within the nozzle.
            
            The specific impulse can thus be calculated from conservation of energy
            \begin{equation}
                I_{\mathrm{sp}} \, g_0 = c = \sqrt{2 \, \left ( h_{\mathrm{chamber}} - h_\mathrm{exit} \right )} 
            \end{equation}
            where the enthalpy $h$ includes the enthalpy of formation.  The NASA polynomial fits to the NIST--JANAF thermochemical data were used for these values \cite{mcbride_nasa_2002}.
            
            Both dissociation and ionization of the hydrogen propellant are considered, with the composition represented by
            \begin{equation}
                \mathrm{H}_2 \Rightarrow \left (1-\beta \right )\mathrm{H}_2 + 2 \left (1-\alpha   \right ) \, \beta \, \mathrm{H} + 2 \, \alpha \, \beta \, \mathrm{H}^+ + 2\,   \alpha \, \beta \, \mathrm{e}^-
            \end{equation}
             where $\alpha$ and $\beta$ quantify the degree of ionization and dissociation of the hydrogen. The value of these parameters can be found by simultaneously solving the equilibrium equations of dissociation and ionization as follows 
            \begin{equation}
                K_{p_\mathrm{dis}}(T) = \frac
                { \left[{\frac{2 \, \beta \, (1- \alpha)}{ 1 + \beta + 2 \, \alpha \, \beta}}  \right]^2}
                { \left[{\frac{1 - \beta}{ 1 + \beta + 2 \, \alpha \, \beta}}  \right]} \left(\frac{P}{P_\mathrm{ref}} \right)
            \end{equation}
            and
            \begin{equation}
                K_{p_\mathrm{ion}}(T) = \frac
                { \left[{\frac{2 \, \alpha \, \beta }{ 1 + \beta + 2 \, \alpha \, \beta}}  \right]^2}
                { \left[{\frac{2 \, \beta \, (1 - \alpha)}{ 1 + \beta + 2 \, \alpha \, \beta}}  \right]} \left(\frac{P}{P_\mathrm{ref}} \right)
            \end{equation}
             where $K_{p_\mathrm{dis}}$ and $K_{p_\mathrm{ion}}$ are the equilibrium constants for the dissociation and ionization reactions, respectively.  The numerical values of the equilibrium constants can be found using the Gibbs function data provided in the NASA database. These coupled non-linear algebraic equations must be solved numerically via an iterative algorithm. Once the equilibrium composition is solved for, the enthalpy of the propellant can be found by summing the product of the mass fraction of each component with its temperature-dependent enthalpy.
            
             In the case of equilibrium flow, the composition continuously varies to maintain equilibrium until the hydrogen is fully expanded to molecular hydrogen, which is assumed to occur at 298.15~K, since the NASA polynomials do not extend below this temperature. In the case of frozen flow, the mass fractions of the composition are held constant throughout the expansion, but the enthalpy of each component at the exit is computed at the final expansion temperature (again, taken at 298.15~K). Finally, the $I_\mathrm{sp}$ can be obtained by normalizing the obtained exit velocity by the gravitational acceleration $g_0$.
             
             \begin{figure}[h]
                \centering
                \begin{subfigure}[b]{\columnwidth}
                    \centering
                    \includegraphics[width=\textwidth]{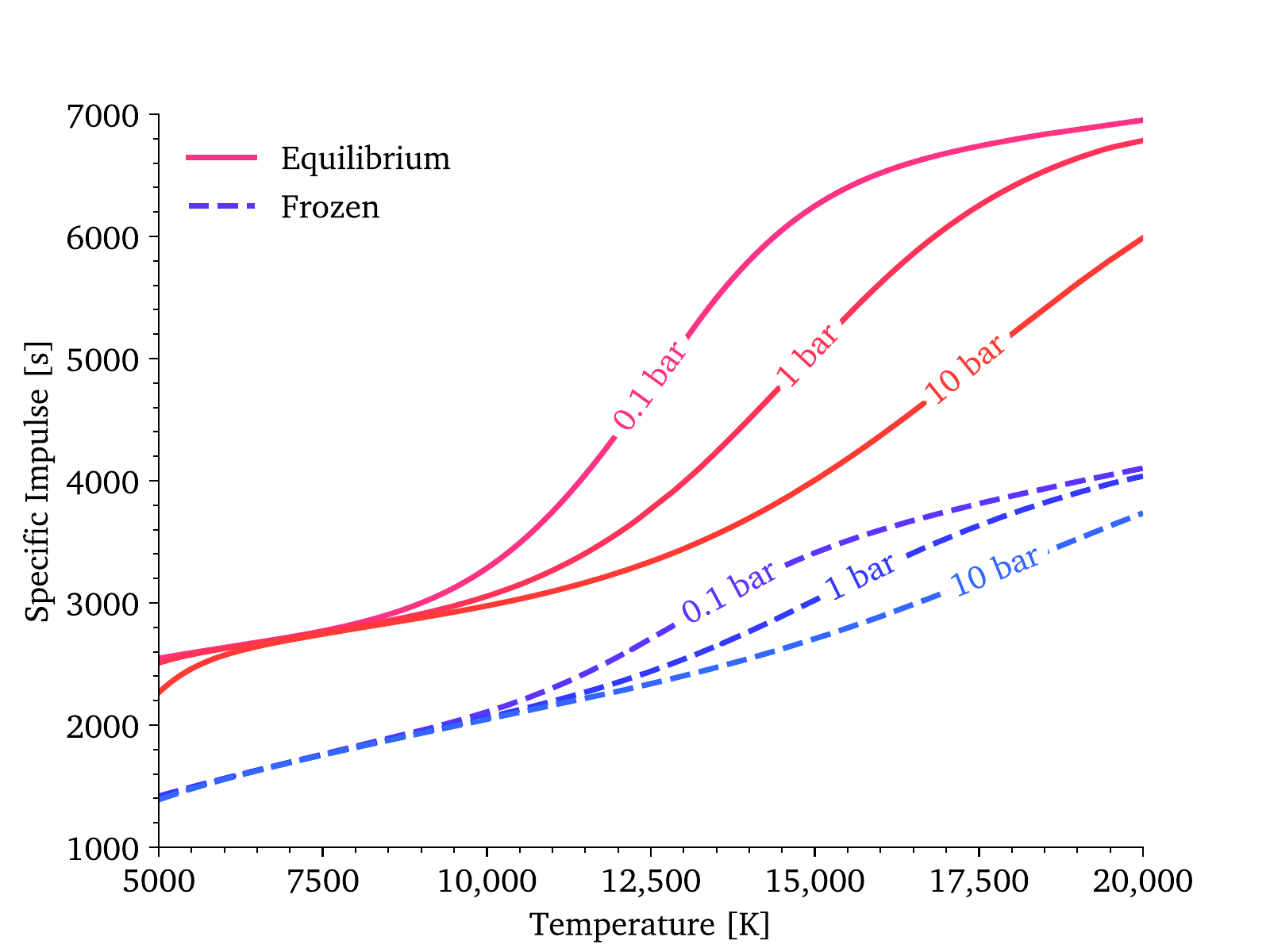}
                    \caption{Effect of temperature}
                    \label{fig:app_IspEqvFrz} 
                \end{subfigure}
                \begin{subfigure}[b]{\columnwidth}
                    \centering
                    \includegraphics[width=\textwidth]{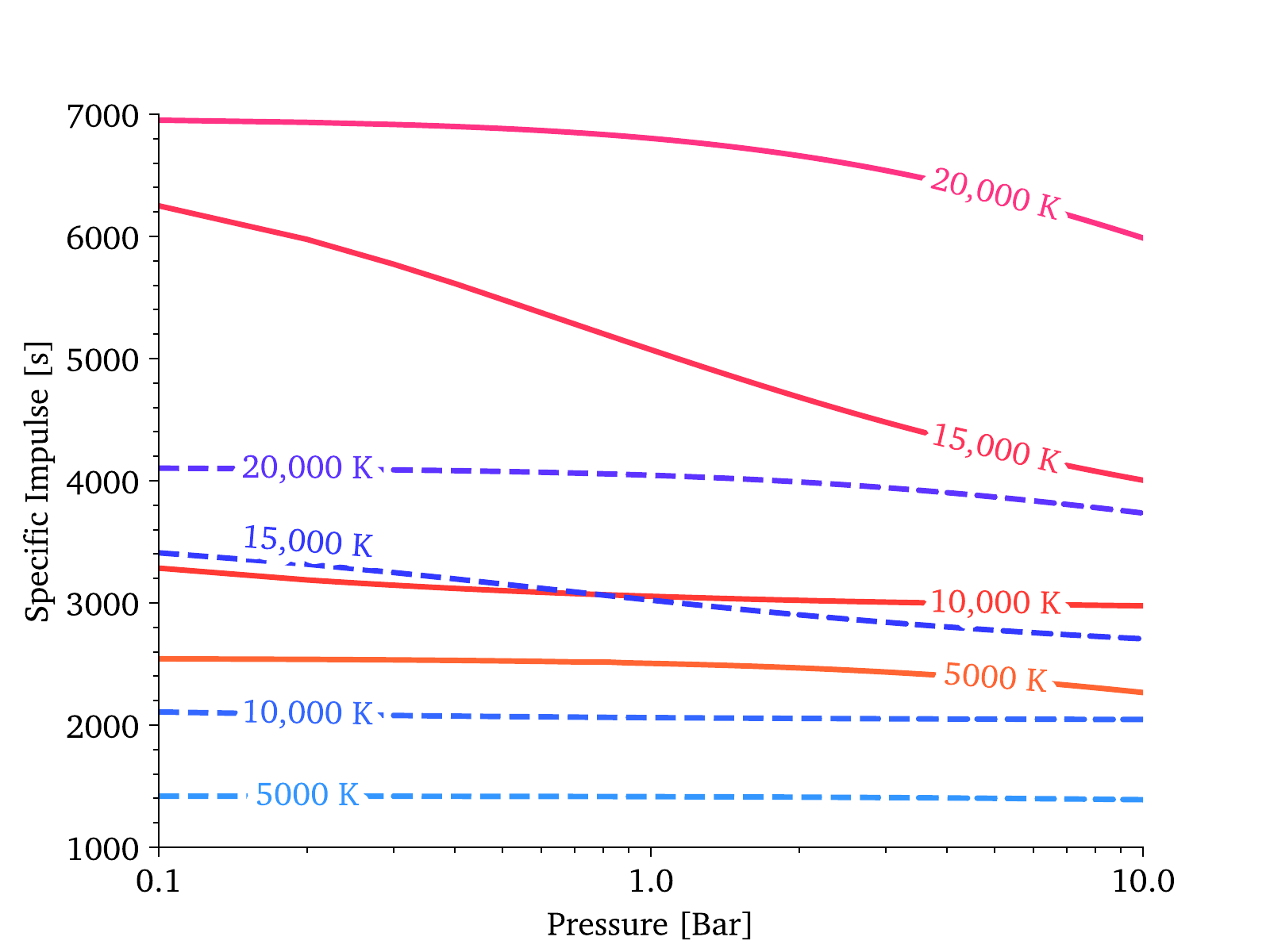}
                    \caption{Effect of chamber pressure}
                    \label{fig:app_IspPressureGrad}
                \end{subfigure}
                \caption{Specific Impulse of laser-heated hydrogen}
                \label{fig:app_IspDependency}
             \end{figure}
             
             The results for specific impulse are plotted in figure \ref{fig:app_IspEqvFrz} as a function of chamber temperature (at various chamber pressures) and chamber pressure (at various chamber temperatures). Note that for chamber temperature values greater than 10,000~K, the equilibrium specific impulse is significantly greater as the chamber pressure is decreased from 10~bar to 0.1~bar, in contrast to the behavior of conventional chemical rockets. The reason for this behavior is that the degree of ionization is greater at lower pressures for the same chamber temperature, and the enthalpy of recombination that occurs as the propellant is expanded in equilibrium results in effectively greater energy being released into the flow, and consequently a greater exhaust velocity is obtained.

        \section{Nomenclature}
            \noindent Symbols:
            \begin{itemize}
                \item[$A$] surface area
                \item[$a$] acceleration
                \item[$C_\mathrm{B}$] ballistic coefficient [kg/m$^2$]
                \item[$c$] exhaust velocity
                \item[$d$] distance or length
                \item[$D$] diameter
                \item[$g$] gravitational acceleration
                \item[$g_0$] standard gravity
                \item[$h$] enthalpy [J/kg]
                \item[$I_\mathrm{sp}$] specific impulse [s]
                \item[$K_p$] chemical equilibrium constant
                \item[$m$] mass
                \item[$P$] pressure [bar]
                \item[$q''$] heat flux [W/m$^2$]
                \item[$R$] radius
                \item[$r$] radial coordinate
                \item[$T$] temperature
                \item[$t$] time
            \end{itemize}
            Greek Symbols:
            \begin{itemize}
                \item[$\alpha$] specific mass [kg/kW]
                \item[$\alpha$] ionized hydrogen mole fraction (\emph{only} in \ref{sec:app_IspCalc})
                \item[$\beta$] dissociated hydrogen mole fraction
                \item[$\epsilon$] emissivity
                \item[$\eta$] efficiency
                \item[$\lambda$] wavelength
            \end{itemize}
            Subscripts:
            \begin{itemize}
                \item[0] initial (wet mass)
                \item[ch] characteristic
                \item[e] laser emitter or array
                \item[f] focal (length) 
                \item[g] glass-transition
                \item[in] inlet of transpiration pore
                \item[m] maneuver
                \item[out] outlet of transpiration pore
                \item[pl] payload
                \item[pr] propellant
                \item[ps] propulsion system
                \item[r]  laser receiver (e.g. reflector or PV array) 
                \item[rad] radiation
            \end{itemize}

\bibliographystyle{elsarticle-num} 
\bibliography{references}





\end{document}